\documentclass{journal}

\usepackage{amsmath, graphicx, latexsym, amssymb,  amsfonts}
\usepackage[mathscr]{eucal}
\usepackage{graphics}
\usepackage{graphicx}
\usepackage{subfig}
\usepackage{epstopdf}
\usepackage{color}
\usepackage{epsfig}
\newcommand{\AddedTh}[1]{#1}

\newcommand{\la}{\lambda}
\newcommand{\eps}{\varepsilon}
\newcommand{\ph}{\varphi}

\newcommand{\al}{\alpha}

\newcommand{\sig}{\sigma}
\newcommand{\del}{\delta}

\newcommand{\Gam}{\mathnormal{\Gamma}}
\newcommand{\Del}{\mathnormal{\Delta}}

\newcommand{\X}{\mathnormal{\Xi}}

\newcommand{\Om}{\mathnormal{\Omega}}

\newcommand{\R}{{\mathbb R}}

\newcommand{\EE}{{\mathbb E}}

\newcommand{\PP}{{\mathbb P}}

\newcommand{\calA}{{\cal A}}
\newcommand{\calB}{{\cal B}}

\newcommand{\calD}{{\cal D}}
\newcommand{\calE}{{\cal E}}
\newcommand{\calF}{{\cal F}}

\newcommand{\calH}{{\cal H}}
\newcommand{\calI}{{\cal I}}

\newcommand{\calL}{{\cal L}}

\newcommand{\calU}{{\cal U}}

\newcommand{\calX}{{\cal X}}

\newcommand{\bB}{{\mathbf B}}

\newcommand{\bT}{{\mathbf T}}

\newcommand{\bx}{{\mathbf x}}

\DeclareMathOperator*{\argmin}{arg\,min}
\DeclareMathOperator*{\argmax}{arg\,max}

\author{Mark Shifrin \\Department of Electrical Engineering\\Technion--Israel Institute of Technology} 
\begin{document}

\title{An asymptotically optimal policy and state-space collapse for the multi-class shared queue} 



\newtheorem{lemma}{Lemma}
\newtheorem{theorem}{Theorem}
\newtheorem{corollary}{Corollary}
\newtheorem{definition}{Definition}
\newtheorem{proposition}{Proposition}

\newtheorem{remark}{Remark}

\def\thelemma{\arabic{section}.\arabic{lemma}}
\def\thetheorem{\arabic{section}.\arabic{theorem}}
\def\thecorollary{\arabic{section}.\arabic{corollary}}
\def\thedefinition{\arabic{section}.\arabic{definition}}
\def\theexample{\arabic{section}.\arabic{example}}
\def\theproposition{\arabic{section}.\arabic{proposition}}
\def\thecondition{\arabic{section}.\arabic{condition}}
\def\theassumption{\arabic{section}.\arabic{assumption}}
\def\theconjecture{\arabic{section}.\arabic{conjecture}}
\def\theproblem{\arabic{section}.\arabic{problem}}
\def\theremark{\arabic{section}.\arabic{remark}}

\newcommand{\manualnames}[1]{
\def\thelemma{#1.\arabic{lemma}}
\def\thetheorem{#1.\arabic{theorem}}
\def\thecorollary{#1.\arabic{corollary}}
\def\thedefinition{#1.\arabic{definition}}
\def\theexample{#1.\arabic{example}}
\def\theproposition{#1.\arabic{proposition}}
\def\theassumption{#1.\arabic{assumption}}
\def\theremark{#1.\arabic{remark}}
}

\newcommand{\beginsec}{
\setcounter{lemma}{0}
\setcounter{theorem}{0}
\setcounter{corollary}{0}
\setcounter{definition}{0}
\setcounter{example}{0}
\setcounter{proposition}{0}
\setcounter{condition}{0}
\setcounter{assumption}{0}
\setcounter{conjecture}{0}
\setcounter{problem}{0}
\setcounter{remark}{0}
}

\newcommand{\s}{\sigma}

\newcommand{\scrA}{\mathscr{A}}
\newcommand{\scrM}{\mathscr{M}}
\newcommand{\scrS}{\mathscr{S}}
\newcommand{\scrI}{\mathscr{I}}

\newcommand{\frA}{\mathfrak{A}}
\newcommand{\frM}{\mathfrak{M}}
\newcommand{\frS}{\mathfrak{S}}

\newcommand{\lan}{\langle}
\newcommand{\ran}{\rangle}
\newcommand{\uu}{\underline}
\newcommand{\oo}{\overline}
\newcommand{\skp}{\vspace{\baselineskip}}
\newcommand{\supp}{{\rm supp}}
\newcommand{\diag}{{\rm diag}}
\newcommand{\trace}{{\rm trace}}
\newcommand{\w}{\wedge}
\newcommand{\lt}{\left}
\newcommand{\rt}{\right}
\newcommand{\pl}{\partial}
\newcommand{\abs}[1]{\lvert#1\rvert}
\newcommand{\norm}[1]{\lVert#1\rVert}
\newcommand{\mean}[1]{\langle#1\rangle}
\newcommand{\To}{\Rightarrow}
\newcommand{\til}{\widetilde}
\newcommand{\wh}{\widehat}
\newcommand{\dist}{{\rm dist}}
\newcommand{\grad}{\nabla}
\newcommand{\iy}{\infty}

\newcommand{\be}{\begin{equation}}
\newcommand{\ee}{\end{equation}}

\newcommand{\tab}{\hspace*{0.3in}}
\newcommand{\Tab}{\hspace*{1.0in}}
\newcommand{\no}{\nonumber}
\newcommand{\noi}{\noindent}
\newcommand{\txt}{\textrm}
\newcommand{\ds}{\displaystyle}
\newcommand{\RR}{\mathbb{R}}
\newcommand{\vf}{\varphi}

\definecolor{co}{rgb}{0.8,0,0.8}
\definecolor{gr}{gray}{0.5}
\newcommand{\gr}{\color{gr}}
\newcommand{\vp}{\varepsilon}

\catcode`\@=11

\catcode`\@=12

\newcommand{\uln}{\underline}


%

\maketitle

\begin{abstract}

We consider a multi-class G/G/1 queue with a finite shared buffer.
There is task admission and server scheduling control which aims to minimize the cost which consists of holding and rejection components. We construct a policy that is asymptotically optimal in the heavy traffic limit.
The policy stems from solution to Harrison-Taksar (HT) free boundary problem and is expressed by  a single free boundary point.
We show that the HT problem solution translated into the queuelength processes follows a specific {\it triangular} form. This form implies the queuelength control policy which is different from the known $c\mu$ priority rule and has a novel structure.

We exemplify that the probabilistic methods we exploit can be successfully applied to solving scheduling and admission problems in cloud computing.



\end{abstract}
\section{Introduction}

We consider the problem of finding asymptotically
optimal (AO) controls for the multiclass G/G/1 queue with a single \textit{shared} buffer, in heavy traffic.
The system is characterized by $I$ classes of arriving tasks. Each task class has its designated queue. The queues share a single buffer, such that the total number of tasks in all queues is limited by its size. We assume that tasks of all classes occupy equally sized storage slots. Upon arrival of a task of class $i$ (with $i\in\{1,\ldots,I\}$),
a decision maker (DM) may either accept or reject it. If the task is admitted it joins the tail of one of the $I$ queues.
In addition, the DM controls
the \textit{fraction of effort} devoted by the \textit{server} to the task at the head of queue $i$,
for each $i$.
Denote holding cost per time unit and rejection cost per customer of class-$i$ as $h_i$ and $r_i$, respectively. Denote reciprocal mean service time by $\mu_i$.
We refer to the two elements of control as {\it admission control} and {\it scheduling control}.
The problem considered is to minimize the combination of holding and rejection costs.
The motivation for this setting is inspired by the results demonstrated in~\cite{Atar-Shif}. However, there is crucial difference which is expressed in the buffer structure.



We assume a critical load condition and observe the model at the diffusion scale.
In the scaling (diffusion) limit, the heavy traffic limits of Queuing control problem (QCP) turn to be a {\it Brownian control problem} (BCP).
It is shown in~\cite{Atar-Shif} that there is an equivalence of an $I$-dimensional BCP
and {\it reduced BCP} (RBCP), where a {\it workload} is a one-dimensional controlled state process.
See references therein for additional discussion on BCP reduction.

The specific one-dimensional RBCP is related to {\it Hamilton-Jacobi-Bellman} (HJB) equation which,  in our setting, takes the form of an ordinary differential equation.
The solution to this problem was analyzed by Harrison and Taksar \cite{har-tak},
as a singular control problem for a Brownian motion (BM), and is given by a reflected BM (RBM), and by a free boundary point of the workload, denoted by $\bf x$. Thus, we are interested in the solution to the one-dimensional \textit{workload control} in the interval of the form $[0,{\bf x}^*]$.

Hence, we treat Harrison-Taksar free boundary problem, specializing in the shared buffer. While the admission control is based on the simple indexing of
$r_i\mu_i$ and the free boundary point ${\bf x}^*$, the scheduling control is complex.
Namely, the difficulty comes from the shared buffer constraint, and is expressed in understanding the structure of the holding cost of the workload process.
The expression for the workload holding cost, denoted by $\bar h$, (equation \eqref{08}), is motivated by the aforementioned equivalence of the BCP and RBCP. 
In particular, $\bar h(w)$, where $w$ is a given workload point, is minimized.
In contrary to the case of dedicated buffers, the solution for the holding cost in case of a shared buffer, is not immediate. Therefore, we formulate and solve this problem as a linear programming (LP) problem.
Additionally, the solution is numerically demonstrated in this work.

Using the solution obtained from the LP and the simple $r\mu$ indexing, we construct a particular AO admission and scheduling policy which is specified by the free boundary point that is used in solving the BCP.
To shortly summarize the policy, 
the domain of the queuelengths in a shared buffer can be geometrically represented by an $I$-simplex in the positive orthant. For example, in $2$-dimensional system, it can be described by a triangle and in the $3$-dimensional system by a tetrahedron. We show that the solution is such that the queuelength process always follows one of the {\it edges} of the simplex. Hence, because of this structure of the queuelength, we merely term it as {\it triangular} policy.

The indexes of $h_i$ and $\mu_i$ are used for scheduling. We make a comparison between the system of dedicated buffers for all task classes, the case which analyzed in~\cite{Atar-Shif} (we name it as a \textit{rectangular} case), and the shared buffer system which is analyzed here, the triangular case. We demonstrate that the priority indexing differs from the one used for the rectangular case and, in particular, from the $c\mu$ priority rule (\cite{BVW}). In contrary, as we analyze in section~\ref{sec4}, the resulting solution shows that there are \textit{at most two classes that can be concurrently present in the shared buffer}.
The key for setting the indexes stems from partitioning of the workload (one-dimensional) scale as follows. First, the buffer is filled up with tasks of the index, which is picked by the lowest $h_i\mu_i$. This is the first index. Next, then the buffer is full and the workload increases, tasks of second index class are accumulated, at expense of tasks of the first one. Once the workload continues to grow, the classes with higher indexes gradually replace those with the lower indexes.
Thus, there are $J\leq I$ workload intervals, such that a unique pair of classes is designated to each interval. The indexes of the classes which fill the buffer are found by the LP.



Following the LP solution at the limit, we formulate the asymptotically optimal policy, such that when the workload level is below ${\bf x}^*$, all arrivals are admitted, with high probability. That is, forced rejections, which occur when the buffer size constraint is reached, have low probability. This is done by assigning dynamic priorities, such that classes which do not belong to the pair of classes which fill the buffer are immediately served. As a result, nearly all rejections occur when the workload exceeds ${\bf x}^*$, and only from one class, which is identified by the lowest $r\mu$ product.
Under the AO policy, the $I$-dimensional queuelength process converges
to the process solving the RBCP. This convergence is a form of a {\it state space collapse} (SSC).
Namely, the queuelength process limits are dictated by the workload process limit.
Note that the full statement of AO is accomplished by proving that the BCP value function is a lower bound on the limit inferior of QCP costs under any sequence of policies. This part is sufficiently general and is not covered in this paper. (See in~\cite{Atar-Shif}, Theorem 1 for the details and the proof).

Various models associated with BCP were treated in~\cite{har1988},~\cite{Har-Van},~\cite{har2000}, \cite{har2003},~\cite{Har-Wil} and others. Examples of characterization of the BCP and RBCP value functions as solutions to HJB equations are found in~\cite{AB2006},~\cite{ABW}.
Scheduling policies, such as $c\mu$ rule and extensions can be also studied from~\cite{BVW} and \cite{van}, and references therein. A general setting of SSC was considered in~\cite{bramson-ssc} and~\cite{williams-ssc}.

We finalize the introduction by bringing the practical motivation for the described setting.
Our primary interest in this problem comes from recent developments in
the application area of cloud computing. In particular, the constantly growing intensity of incoming, outgoing and traversing traffic in the public cloud (e.g. Amazon) makes the diffusion approximation, which we use as our main analytical tool, rather practical.
We bring two concrete practical examples. First, consider a hybrid cloud,
where a private cloud (namely, a \textit{local server}) has a given capacity and
memory limits. The tasks are served by sharing the computing effort among various task types. Once the incoming tasks cannot be queued for the local processing, they are rejected from
a private cloud and sent to a public cloud, where a fixed charge per usage applies.
In this context,~\cite{Atar-Shif} only treated the case where locally processed tasks compete for the server, having dedicated storage resources.
In our case, the problem introduces additional motivation; namely,\textit{ tasks of different types compete with each other over the storage resources} as well. In this work, the common storage is modeled by a shared buffer.
The second example refers to virtual machines (VM) allocation, a procedure which is performed by a hypervisor \textit{residing on the side of the cloud}'s hardware. The VM allocation is based on a common computational resource, and their total number is naturally limited by hardware constraints. Since VM for various types of tasks are instantiated by a single hypervisor, one can view this resource as a buffer. Different types of tasks have different communication and computation demands which are expressed in differentiated service effort. (Assume the overhead and order of VM release and allocation are embodied  in the service rate.) The rejection stands for inability to allocate an additional VM, while the rejection price stands for the VM migration cost.
For further details on modeling of hybrid cloud applications, control and additional practical examples for the model analyzed in this work the reader is referred to \cite{shi-ata-cid},~\cite{shifrin2013thesis} and references therein.

{\it Notation}

We will use the following notation. Given $k\in\mathbb N$, $\{e^{(i)},i=1,\ldots,k\}$ denote the
standard basis in $\R^k$.
For $x\in\R$, $x^+=\max(x,0)$. For $a,b\in\R^k$, $a=(a_i)_{i=1,\ldots,k}$, $b=(b_i)_{i=1,\ldots,k}$,
we denote $\|a\|=\sum_{i=1}^k|a_i|$ and $a\cdot b=\sum_{i=1}^ka_ib_i$.
For $y:\R_+\to\R^k$ and $T>0$, $\|y\|_T=\sup_{t\in[0,T]}\|y(t)\|$.
The modulus of continuity of $y$ is given by
\[
\bar w_T(y;\theta)=\sup\{\|y(s)-y(t)\|:s,t\in[0,T],|s-t|\le\theta\}, \qquad \theta,T>0.
\]
Denote $A[s,t]=A(t)-A(s)$ for any process $A$.

The structure of the rest of the paper is as follows. In sections~\ref{sec2} and ~\ref{sec22} we bring the definitions, which, in most parts, closely follow the definitions from~\cite{Atar-Shif}. We start with the queuing and diffusion models.
We next define the Brownian control problem (BCP), and the reduced Brownian control problem (RBCP).  Proposition~\ref{prop1} shows the relation between these two problems. These components are consistent with~\cite{Atar-Shif}, and we bring them here because we will use them in the proof in section~\ref{sec4}. We analyze HT solution for the shared buffer and bring a detailed numerical example in section~\ref{sec:har-tak}.
Section~\ref{sec4} constitutes the formulation of nearly optimal scheduling and admission policy. In Section~\ref{sec:ssc} we state and prove Theorem~\ref{th2}, which, altogether with the lower bound stated in~\cite{Atar-Shif}, implies the AO of the proposed policy. We conclude in Section~\ref{sec:concl}.

\section{Queueing and diffusion models}
\label{sec2}
We start with definitions of queueing and diffusion models. 
Consider a sequence of systems, indexed by a superscript $n\in\mathbb N$. 
The system has a single server and single shared buffer, where
tasks of all classes can occupy exactly one slot per task.
The capacity of the buffer is limited. 
Customers that arrive at the system are judged by the DM to either be accepted or rejected.
Those that are accepted are queued in the common queue. 
Within each class, service is provided in the order of arrival. 
Processor sharing is allowed, in the sense that the server
is capable of serving up to $I$ customers (of distinct classes) simultaneously, but the service cannot be shared within the customers of the same class.
Denote an {\it allocation vector} of size $I$, representing the fractions
of effort dedicated to the classes, and is chosen from the set
\[
\calB:=\Big\{\beta\in\R_+^I: \sum_{i\in\calI}\beta_i\le1\Big\},
\]
where, $\calI=\{1,2,\ldots,I\}$.

We assume a given probability space $(\Om,\calF,\PP)$ with expectation w.r.t.\ $\PP$ denoted by $\EE$.
Arrivals occur according to independent renewal processes.
The parameters $\la^n_i >0$, $i\in\calI$, $n\in\mathbb N$, satisfying
$\la^n_i=n\la_i+\sqrt n\hat\la_i+o(\sqrt n)$, with fixed positive $\lambda_i$ and $\hat\la_i\in\R$,
represent the {\it reciprocal mean
inter-arrival times} of class-$i$ tasks in the $n$-th system.
The parameters  $\mu^n_i>0$, $i\in\calI$, $n\in\mathbb N$, satisfying
$\mu^n_i=n\mu_i+\sqrt n\hat\mu_i+o(\sqrt n)$, with fixed positive $\mu_i$ and $\hat\mu_i\in\R$,
represent the {\it reciprocal mean
service times} of class-$i$ tasks in the $n$-th system.
Let $\{{\it IA}_i(l) : l\in\mathbb N\}_{i\in\calI}$ be independent sequences
of strictly positive i.i.d.\ random variables with mean
$\mathbb E[{\it IA}_i(1)]=1$, $i\in\calI$ and squared coefficient of variation
${\rm Var}({\it IA}_i(1))/\mathbb E[{\it IA}_i(1)^2]=C^2_{{\it IA}_i}\in(0,\iy)$.
The number of arrivals of class-$i$ customers up to time $t$, for the $n$-th system,
is given by
\begin{equation}\label{28}
A^n_i(t)=A_i(\la^n_it), \quad \text{ where }\quad
A_i(t)=\sup\Big\{l\geq 0:\sum_{k=1}^l{\it IA}_i(k)\le t\Big\},\quad t\geq 0.
\end{equation}
Let independent sequences $\{{\it ST}_i(l) : l\in\mathbb N\}_{i\in\calI}$ of strictly
positive i.i.d.\ random
variables be given, with mean $\mathbb E[{\it ST}_i(1)]=1$ and squared coefficient of variation
${\rm Var}({\it ST}_i(1))/\mathbb E[{\it ST}_i(1)^2]=C^2_{{\it ST}_i}\in(0,\iy)$.
The time required to complete the $l$-th service to
a class-$i$ customer in the $n$-th system is given
by ${\it ST}_i(l)/\la_i^n$ units of time dedicated by the server to
this class. This is also known as
{\it potential service time} processes, that is
\begin{equation}\label{12}
S^n_i(t)=S_i(\mu^n_it), \quad \text{ where }\quad
S_i(t)=\sup\Big\{ l\geq 0:\sum_{k=1}^l{\it ST}_i(k)\leq t\Big\},
\quad t\geq 0.
\end{equation}
$S^n_i(t)$ is the number of class-$i$ jobs completed by the time when the server has dedicated
$t$ units of time to work on jobs of this class.
We assume the sequences $\{{\it IA}_i\}$ and $\{{\it ST}_i\}$ are independent.
We assume the {\it critical
load condition}
\begin{equation}
  \label{15}
  \sum_{i\in\calI}\rho_i=1,\qquad \text{where}\qquad \rho_i=\frac{\la_i}{\mu_i},\,i\in\calI.
\end{equation}

The number of class-$i$ rejections until time $t$ and customers present at time $t$ in the $n$-th system
 is denoted by $Z^n_i(t)$ and $X^n_i(t)$, correspondingly.
Since rejections occur only at times of arrival, we have
\begin{equation}\label{99}
Z^n_i(t)=\int_{[0,t]}z^{n,i}_sdA^n_i(s)
\end{equation}
for some process $z^{n,i}$.

We call $X^n=(X^n_i)_{i\in\calI}$ the queuelength process.
We assume, deterministic $X^n_i(0)$ and that no partial service has been provided to
any of the tasks present in the system at time zero.
Let $B^n=(B^n_i)_{i\in\calI}$
be a process taking values in the set $\calB$. 
Then
\be\label{2}
T^n_i(t)=\int_0^t B^n_i(s)ds
\ee
gives the time devoted to class-$i$ customers up to time $t$.
The number of service completions of class-$i$ jobs during
the time interval $[0,t]$ is given by 
\be\label{1}
D^n_i(t)=S^n_i(T^n_i(t)).
\ee
We thus have
\be\label{3}
X^n_i(t)=X^n_i(0)+A^n_i(t)-D^n_i(t)-Z^n_i(t)=X^n_i(0)+A^n_i(t)-S^n_i(T^n_i(t))-Z^n_i(t),\qquad t\ge0.
\ee
It is assumed that $B^n,S^n$, $Z^n$, $D^n$, $X^n$, $T^n$ have RCLL sample paths.
Next define a rescaled version of the processes at diffusion
scale as
\begin{equation}\label{16-}
\hat A^n_i(t)=\frac{A^n_i(t)-\la^n_it}{\sqrt n},\qquad
\hat S^n_i(t)=\frac{S^n_i(t)-\mu^n_it}{\sqrt n},\qquad i\in\calI,
\end{equation}
\[
\hat Z^n(t)=\frac{Z^n(t)}{\sqrt n},\qquad
\hat X^n(t)=\frac{X^n(t)}{\sqrt n}.
\]
The shared buffer structure is specified as follows
\[
\calX=\{y\in\R_+^I:\sum_iy_i\le b\}
\]
We assume that the rescaled initial condition $\hat X^n(0)$ also lies in $\calX$.
The buffer constraint is always met, namely:
\begin{equation}
  \label{16}
  \hat X^n(t)\in\calX,\qquad t\ge0,\, a.s.
\end{equation}
The rejection mechanism assures the condition above by rejecting arrivals occurring at a time $t$ when
$(X^n(t-)+e^{(i)})/\sqrt n\not\in\calX$.
We refer to these rejections as
{\it forced rejections}.
In our setting, we distinguish them from admission/rejection decisions which are part of the control process. Note that the actual un-normalized buffer size scales like $\sqrt n$.

The control process $U^n=(Z^n,B^n)$, which is determined based on observations
from the past (and present) events in the system, is defined as follows.
\begin{definition}\label{def1} {\bf (Admissible control, QCP)}
Fix $n\in\mathbb N$ and consider fixed processes $(A^n,S^n)$ given by
\eqref{28} and \eqref{12}. 
A process $U^n=(Z^n,B^n)$, taking values in $\R_+^I\times\calB$, having RCLL
sample paths with the processes $Z^n_i$, $i\in\calI$ having nondecreasing sample paths
and given in the form \eqref{99},
is said to be an admissible control for the $n$-th system if the following holds.
Let the processes $T^n$, $D^n$, $X^n$ be defined by the $A^n$ and $S^n$ and control
processes, $(A^n,S^n)$ and $(Z^n,B^n)$,
via equations \eqref{2}, \eqref{1} and \eqref{3}, respectively.
Then
\begin{itemize}
\item $(Z^n,B^n)$ is adapted to the filtration
\(
 \sig\{A^n_i(s),D^n_i(s), i\in\calI, s\leq t\};
\)
\item One has a.s., that, for all $i\in\calI$ and $t\ge0$,
\be
X^n_i(t)=0 \quad \text{implies} \quad B^n_i(t)=0.
\label{4}
\ee
\end{itemize}
An admissible control under which the scaled version $\hat X^n$ of $X^n$ satisfies \eqref{16}
is said to satisfy the buffer constraint.
\end{definition}
We denote the class of all admissible controls $U^n=(Z^n,B^n)$ satisfying the buffer constraints,
by $\calU^n$. 
Fix $\al>0$, $h\in(0,\iy)^I$ and $r\in(0,\iy)^I$. For each $n\in\mathbb N$ consider the cost
\be\label{5}
J^n(U^n)=\mathbb E\Big[\int_0^\iy e^{-\al t}[h\cdot\hat X^n(t)dt+r\cdot d\hat Z^n(t)]\Big] 
\ee
The QCP value is given by
\begin{equation}\label{22}
V^n=\inf_{U^n\in\calU^n}J^n(U^n).
\end{equation}
Denote by $\theta^n=(\theta^n_i)_{i\in\calI}$, $\theta^n_i=1/\mu^n_i$,
and $\theta=(\theta_i)_{i\in\calI}$, $\theta_i=1/\mu_i$.
The process $\theta^n\cdot X^n$, referred to as {\it workload}, its
normalized version $\theta^n\cdot\hat X^n$ and its formal limit,
$\theta\cdot X$, will take part in the state-space collapse. 

\section{The Brownian control problems}
\label{sec22}
We address now the \textit{limit problems}.
First, we show that the scaled processes defined in the previous section give rise to the limit processes of the queuelengths. Hence, the definition of the $I$-dimensional BCP of the queuelengths follows. Next, we define the one-dimensional BCP of the workload. Finally, we show the \textit{equivalence of the value functions} of these two problems.

Using \eqref{15}, \eqref{3} and the definition of the rescaled processes, the following identity holds for $i\in\calI$ and $t\ge0$:
\begin{equation}
  \label{17}
  \hat X^n_i(t)=\hat X^n_i(0)+\hat W^n_i(t)
  +\hat Y^n_i(t)-\hat Z^n_i(t),
\end{equation}
where, denoting $m_i=\hat\la_i-\rho_i\hat\mu_i$, $ m^n_i=\frac{\la^n_i-\rho_i\mu^n_i}{\sqrt n}=m_i+o(1)$,
\begin{equation}
  \label{26}
  \hat W^n_i(t)=\hat A^n_i(t)-\hat S^n_i(T^n_i(t))+m^n_it,
\end{equation}
and
\begin{equation}
  \label{18}
  \hat Y^n_i(t)=\frac{\mu^n_i}{\sqrt n}(\rho_it-T^n_i(t)).
\end{equation}
Since $\sum_i\rho_i=1$ and one always has $\sum_iB^n_i(t)\le1$, it follows that
\begin{equation}\label{21}
\theta^n\cdot\hat Y^n \quad \text{is a nonnegative, nondecreasing process}.
\end{equation}

We are interested in limit problem, where the diffusion coefficient $n$ is taken to infinity. As far as the real system is concerned, we view configuration where task arrival rate and, consequently, the service rate grow large. This connection makes the following model to be of a particular practical interest.
Formally, the limits of \eqref{17}--\eqref{21} and \eqref{5}--\eqref{22} give rise to a control problem associated with diffusion. 
Consider equation \eqref{17}. We assume that the scaled initial conditions $\hat X^n(0)$ converge to $x$ as $n\to\iy$. 
Next,
the centered, rescaled renewal process $\hat A^n_i$ [resp., $\hat S^n_i$]
converges weakly to
a BM starting from zero, with zero mean and diffusion coefficient $\sqrt{\la_i}C_{{\it IA}^i}$
[resp., $\sqrt{\mu_i}C_{{\it ST}^i}$] (see Section 17 of \cite{Bill}).
Assume the processes involved in \eqref{17} give rise
to a limiting BCP.
Then it follows $\hat Y^n$ in \eqref{18} are order one as $n\to\iy$ and, consequently, $T^n(t)$ converge to $\rho t$. Thus, applying the time change in the second term of r.h.s. of \eqref{26}, one sees that $\hat W^n$ converges to $(m,\sig)$-BM starting from zero, with drift vector $m=(m_i)_{i\in\calI}$
and diffusion matrix $\sig=\diag(\sig_i)$, where
\[
\sig_i^2:=\la_iC^2_{{\it IA}^i}+\mu_iC^2_{{\it ST}^i}\rho_i
=\la_i(C^2_{{\it IA}^i}+C^2_{{\it ST}^i}).
\]
As for $\hat Y^n$, it gives rise to a process $Y$ such that $\theta\cdot Y$
is nonnegative and nondecreasing. The process $\hat Z^n$ gives rise to a process $\theta\cdot Z$
with nonnegative, nondecreasing components.
\subsubsection*{\sl The BCP}

\begin{definition}\label{def2} {\bf (Admissible control, BCP)}
An admissible control for the initial condition $x_0\in\calX$
is a filtered probability space $(\Om',\calF',\{\calF'_t\},\PP')$ for which
there exists an $(m,\sig)$-BM, $W$, and positive process $U=(Y,Z)$, with RCLL sample paths, such that the following conditions hold:
\begin{itemize}
\item
$W$, $Y$ and $Z$ are adapted to $\{\calF'_t\}$;
\item
\begin{equation}\label{29}
\text{For $0\le s<t$, the increment $W_t-W_s$ is independent of $\calF'_s$ under $\PP'$;}
\end{equation}
\item
\begin{equation}
  \label{03}
  \text{$\theta\cdot Y$ and $Z_i$, $i=1,\ldots,I$, are nondecreasing;}
\end{equation}
\item With
\begin{equation}
  \label{02}
  X(t)=x_0+W(t)+Y(t)-Z(t),\qquad t\ge0,
\end{equation}
one has
\begin{equation}
  \label{01}
  X(t)\in\calX\qquad \text{for all $t$, $\PP'$-a.s.}
\end{equation}
\end{itemize}
\end{definition}
In what follows we will always assume that $(Y,Z)$ are admissible controls. Denote the class of such controls as $\calA(x_0)$, where $x_0$ stands for the initial condition.
Let
\begin{equation}
  \label{04}
  J(x_0,Y,Z)=\EE\Big[\int_0^\iy e^{-\al t}[h\cdot X_tdt+r\cdot dZ(t)]\Big].
\end{equation}
The BCP is to find $(Y,Z)$ that minimize $J(Y,Z)$ and achieve the value
\begin{equation}\label{24}
V(x_0)=\inf_{(Y,Z)\in\calA(x_0)}J(x,Y,Z).
\end{equation}

\subsubsection*{\sl The RBCP\\}
\skp
\indent
The reduced one-dimensional problem is obtained as follows.
Multiply equation \eqref{02} and the processes involved in it
by $\theta$. 
Denote $\bar x_0=\theta\cdot x_0$, $\bar m=\theta\cdot m$ and
\(
\bar\sig^2=\sum\theta_i^2\sig_i^2.
\)
Let
\begin{equation}\label{42}
{\bf x}=\max\{\theta\cdot\xi:\xi\in\calX\}.
\end{equation}
\begin{definition}\label{def3} {\bf (Admissible control, RBCP)}
An admissible control for the initial condition $\bar x_0\in[0,\bx]$
is a filtered probability space $(\Om',\calF',\{\calF'_t\},\PP')$ for which
there exist an $(\bar m,\bar\sig)$-BM, $\bar W$, and a positive process $\bar U=(\bar Y,\bar Z)$
with RCLL sample paths, such that the following
conditions hold:
\begin{itemize}
\item
$\bar W$, $\bar Y$ and $\bar Z$ are adapted to $\{\calF'_t\}$;
\item
For $0\le s<t$, the increment $\bar W_t-\bar W_s$
is independent of $\calF'_s$ under $\PP'$;
\item
\begin{equation}
  \label{07}
  \text{$\bar Y$ and $\bar Z$ are nondecreasing;}
\end{equation}
\item With
\begin{equation}
  \label{06}
  \bar X(t)=\bar x_0+\bar W(t)+\bar Y(t)-\bar Z(t),\qquad t\ge0,
\end{equation}
one has
\begin{equation}
  \label{05}
  \bar X(t)\in[0,\bx]\qquad \text{for all $t$, $\PP'$-a.s.}
\end{equation}
\end{itemize}
\end{definition}
We write $\bar\calA(\bar x_0)$ for the class of admissible controls for the initial condition
$\bar x_0$. Given $(\bar Y,\bar Z)\in\bar\calA(\bar x_0)$, let
\begin{equation}
  \label{08}
  \bar J(\bar x_0,\bar Y,\bar Z)=
  \EE\Big[\int_0^\iy e^{-\al t}[\bar h(\bar X_t)dt+\bar r d\bar Z(t)]\Big],
\end{equation}
where the holding cost and the rejection cost for the one-dimensional problem are defined as follows:
\[
\bar h(w)=\min\{h\cdot\xi:\xi\in\calX, \theta\cdot\xi=w\},\qquad w\in[0,{\bf x}],
\]
\[
\bar r=\min\{r\cdot z:z\in\R_+^I,\theta\cdot z=1\}.
\]
One interprets $\bar r$ as a minimal rejection penalty per unit of work.
Note that $\bar h$ is convex by convexity of the set $\calX$.
Note also that as members of $(0,\iy)^I$, $\theta$ and $h$ cannot be orthogonal,
thus $\bar h(w)>0$ for any $w>0$. Since $\bar h(0)=0$, it follows that $\bar h$ is
strictly increasing. Let
\[
\bar V(\bar x_0)=\inf_{(\bar Y,\bar Z)\in\bar\calA(\bar x_0)}\bar J(\bar x_0,\bar Y,\bar Z).
\]

The following definitions will relate the two problems as follows.
See, that the extremal points of the set $\{z\in\R_+^I:\theta\cdot z=1\}$ are precisely
$\theta_i^{-1}e^{(i)}$, namely $\mu_ie^{(i)}$, $i\in\calI$. Hence there exists
(at least one) $i^*$ such that $\zeta=\mu_{i^*}e^{(i^*)}$ satisfies
\[
\zeta\in\argmin_z\{r\cdot z:z\in\R_+^I,\theta\cdot z=1\}.
\]
Fix such $i^*$ and the corresponding $\zeta$.
See that $i^*$ can be expressed via
\begin{equation}
  \label{69}
  r_{i^*}\mu_{i^*}=\min_ir_i\mu_i.
\end{equation}

Next, let $\gamma:[0,\bx]\to\calX$ be Borel measurable, satisfying
\begin{equation}\label{36}
\gamma(w)\in\argmin_\xi\{h\cdot\xi:\xi\in\calX,\theta\cdot\xi=w\},\qquad w\in[0,{\bf x}].
\end{equation}
By definition,
$\gamma(w)\in\calX$, $\theta\cdot\gamma(w)=w$, and $h\cdot\gamma(w)=\bar h(w)
\le h\cdot\xi$ for every
$\xi\in\calX$ for which $\theta\cdot\xi=w$.
Hence, as it will become clear soon, these definitions of the costs will imply the equivalence of the value functions of BCP and RBCP.
\begin{remark}
One observes that the rejection process in the problem of dedicated buffers (denote it as a \textit{rectangular case}) and that of a shared buffer (which we term as a \textit{triangular case}) will follow the same rule.
However, as opposed to~\cite{Atar-Shif}, the solution for $\bar h(w)$ in the case of shared buffer is not immediately understood. We analyze this in the sequel of this section.
\end{remark}

We now state the key proposition which determines the relation between the BCP and RBCP. This proposition, together with proposition~\ref{prop2} stated in sequel, leads to the formulation of the AO control policy.
\begin{proposition}
  \label{prop1}
  Let $x_0\in\calX$ and $\bar x_0=\theta\cdot x_0$.
\\
  i. Given an admissible control $(\Om',\calF',\{\calF'_t\},\PP',W,Y,Z)$ for $x$
  for the (multidimensional) BCP, define $(\bar W,\bar X,\bar Y,\bar Z)$ by
  $(\theta\cdot W,\theta\cdot X,\theta\cdot Y,\theta\cdot Z)$.
  Then $(\bar Y,\bar Z)\in\bar\calA(\bar x_0)$ and $\bar J(\bar x_0,\bar Y,\bar Z)\le
  J(x_0,Y,Z)$.
\\
  ii. Conversely,
  let an admissible control $(\Om',\calF',\{\calF'_t\},\PP',\bar W,\bar Y,\bar Z)$
  for $\bar x_0$
  for the RBCP be given, and assume the probability
  space supports an $(m,\sig)$-BM $W$. Assume $W$ is $\{\calF'_t\}$-adapted
  and satisfies $\theta\cdot W=\bar W$ and \eqref{29}. Construct $(X,Y,Z)$ by
  \begin{equation}\label{71}
  X(t)=\gamma(\bar X(t)),\qquad Z(t)=\zeta\bar Z(t),
  \end{equation}
  \begin{equation}\label{72}
  Y(t)=X(t)-x_0-W(t)+Z(t).
  \end{equation}
  Then $(Y,Z)\in\calA(x_0)$, and
  $J(x_0,Y,Z)\le \bar J(\bar x_0,\bar Y,\bar Z)$.
\\
  iii. Consequently, $V(x_0)=\bar V(\bar x_0)$.
\end{proposition}

Note that the proposition holds for any convex $\bar h$. Hence, we skip the proof, which the interested reader can find in~\cite{Atar-Shif}.

\section{The Harrison-Taksar free boundary problem} \label{sec:har-tak}

We now focus on the one-dimensional problem. The function $\bar V$ is $C^2[0,\bx]$ and solves the following Bellman equation
\begin{equation}
  \label{25}
  \begin{cases}
  \ds
 \Big[\frac{1}{2}\bar\sig^2f''+\bar mf'-\al f+\bar h\Big]\w f'\w[\bar r-f']=0,
 \qquad \text{in } (0,\bx),
 \\ \\
 f'(0)=0,\qquad
 f'(\bx)=\bar r.
 \end{cases}
\end{equation}
This result has been demonstrated by Harrison and Taksar \cite{har-tak}.
They showed that an optimal control is such that the process $\bar X$ is a RBM. 
We introduce a {\it Skorohod map}, next. Let $a>0$. The Skohorod map on the interval $[a,b]$,
denoted by $\Gam_{[a,b]}$, is map $D([0,\iy):\R)\to D([0,\iy):\R)^3$.
It solves {\it Skorohod Problem}, and its solution is a map 
 $\psi\to(\ph,\eta_1,\eta_2)$, 
for a given $\psi$, a triplet $(\ph,\eta_1,\eta_2)$, such that
\begin{equation}\label{44}
\ph=\psi+\eta_1-\eta_2,\qquad \ph(t)\in[a,b] \text{ for all } t,
\end{equation}
\begin{equation}\label{45}
\text{$\eta_i$ are nonnegative and nondecreasing,
and $\int_{[0,\iy)}1_{(a,b]}(\ph)d\eta_1=\int_{[0,\iy)}1_{[a,b)}(\ph)d\eta_2=0$.}
\end{equation}

See \cite{KLRS} for existence and uniqueness of solutions,
and continuity and further properties of the map.
In particular, it is well-known that $\Gam_{[a,b]}$ is continuous in the uniformly-on-compacts
topology.

The following proposition is mostly a result of \cite{har-tak}.
\begin{proposition}
  \label{prop2}
  The function $\bar V$ is $C^2$ on $[0,\bx]$ and solves \eqref{25} uniquely among
  all $C^2$ functions. 
  Denote $\bx^*=\inf\{y\in[0,\bx]:\bar V'(z)=\bar r \text{ for } z\in[y,\bx]\}$.
  Then $\bx^*\in(0,\bx)$. Fix $\bar x_0\in[0,\bx]$.
  Let $\bar W$ be an $(\bar m,\bar\sig)$-BM and let $\bar X$, $\bar Y$ and $\bar Z$ be
  the corresponding RBM on $[0,\bx^*]$ and boundary terms for $0$ and $\bx^*$, defined as
  \begin{equation}\label{73}
  (\bar X,\bar Y,\bar Z)=\Gam|_{[0,\bx^*]}(\bar x_0+\bar W).
  \end{equation}
  Then $(\bar Y,\bar Z)$ is optimal for $\bar V(\bar x_0)$, i.e.,
  $\bar J(\bar x_0,\bar Y,\bar Z)=\bar V(\bar x_0)$.
\end{proposition}


The reader is referred to~\cite{Atar-Shif} for the proof. An optimal control for the BCP stems from propositions~\ref{prop1} and~\ref{prop2}.

\subsection{Solution for $\bar h$} 

As mentioned, the specifics imposed by the shared property of a buffer implies non-trivial solution to the one-dimensional holding cost. The structure of the workload process is such that
$\bar X=\theta\cdot X$ is given as a RBM on $[0,\bx^*]$, where the free boundary point $\bx^*$
is dictated by the Bellman equation.
The multidimensional queuelength process $X$ is recovered from $\bar X$ by
\(
X=\gamma(\bar X).
\)

Recall that the shared buffer domain has the form
\begin{equation}\label{10}
\calX=\{x\in\R^{I}_+:0\le\sum_{i=1}^{I}x_i\le b, i\in\calI\},
\end{equation}
for some fixed $b>0$.  
As mentioned, there is no difference between the cases of the shared and the dedicated models, in rejection structure.
Namely, the BCP solution implies that in the scaled model, rejections should occur only when the scaled workload exceeds the level $\bx^*$, and only from class $i^*$, for which $r_i\mu_i$ is minimal.
Next, the relation
\begin{equation}\label{90}
\hat X^n=\gamma(\theta\cdot\hat X^n)+o(1)
\end{equation}
between the queuelength and workload processes must hold.
We solve for the minimizing curve $\gamma$, where $\calX$ takes the form \eqref{10}.
Equation \eqref{36} can be written as
\[
\gamma(w)\in\argmin_x\{h\cdot x:0\leq x_i,\;0\le\sum_1^Ix_i\le b\text{ and } \theta\cdot x=w\},
\quad w\in[0,\bx].
\]
To compare a rectangular system and a triangular system, one sees that the description of $\bar h$ in the latter case is not trivial. This is because the minimization is done under restriction of the shared buffer. Namely, in rectangular case, it is clear that as long as the workload grows, the buffers with lower priorities are filled up until their limits are reached. However, in triangular case, one have to explicitly solve for $\bar h$ in order to understand which task types are being preferred for each workload point.

\begin{figure}[h]
\begin{center}
\vspace{1.3in}
\raisebox{-2.2pt}[0pt][10pt]{\Large%
  \textbf{%
    \raisebox{0.3ex}{\includegraphics[width=35mm]{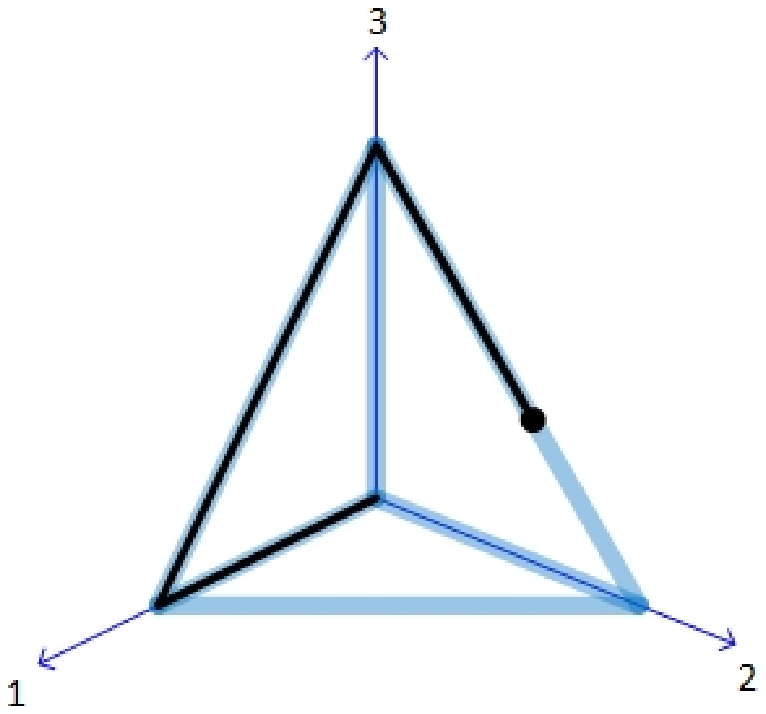}} \hspace{0.6em}
    \raisebox{0.3ex}{\includegraphics[width=35mm]{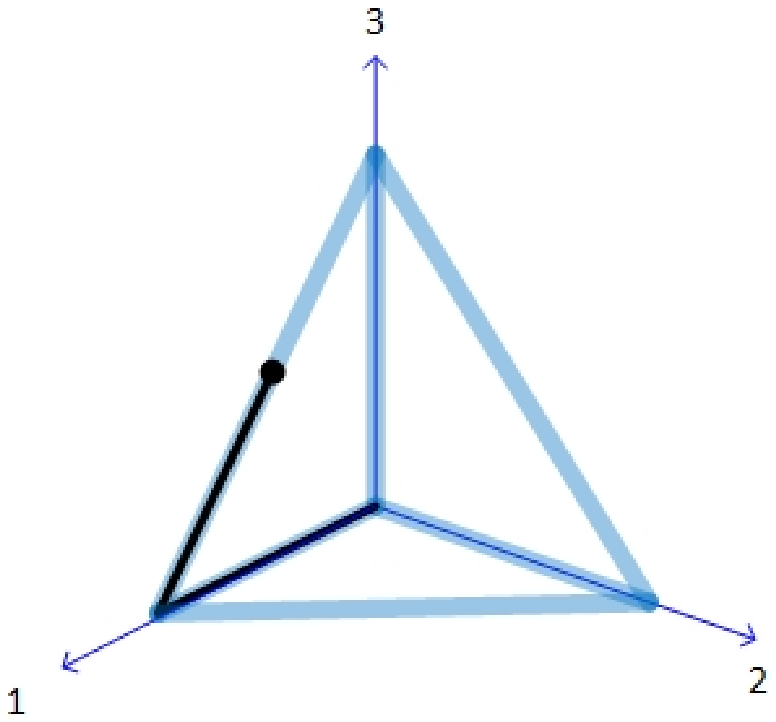}}\hspace{0.6em}
    \raisebox{1.5ex}{\includegraphics[width=25mm]{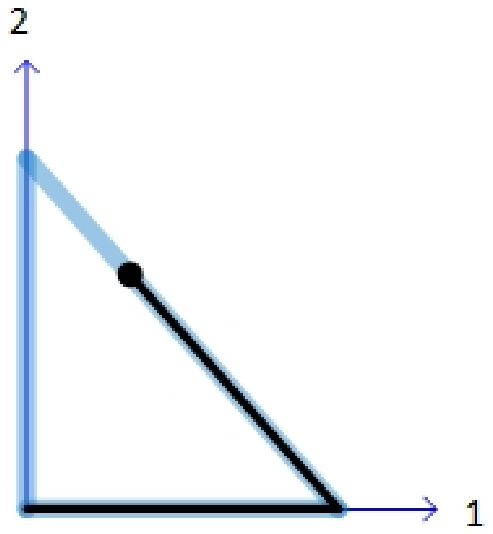}}\hspace{0.6em}
    \raisebox{2.2ex}{\includegraphics[width=25mm]{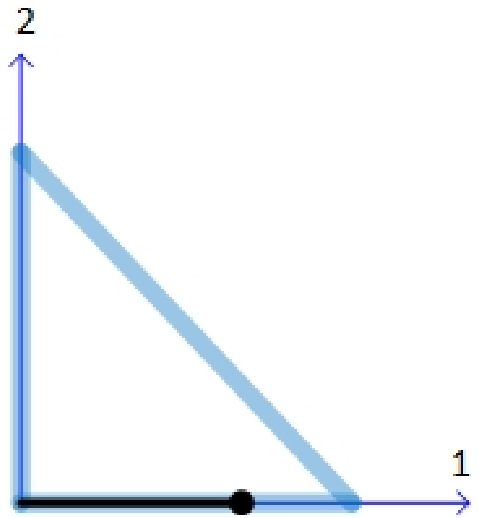}}
  }
}
\caption{\small
The case of 2 sharing classes (right), and 3 sharing classes (left). Rejection point may be reached before the buffer with least priority is admitted. For a higher rejection level, the curve continues
along the boundary.}\label{fig3}
\end{center}
\end{figure}

Hence, we find $\bar h$ by the linear program (LP).
Assume first the buffer is full, and the total workload is $w$. We later extend the solution for the workload points where the buffer is partially empty.
The standard LP form is written
{\small
\begin{align}
& \text{minimize }\sum_{i=1}^Ih_ix_i \;\;\text{   w.r.t } \nonumber\\
& \sum_{i=1}^I\theta_ix_i=w,\qquad \;\sum_{i=1}^Ix_i=b \label{1:lin_std}
\end{align}}
Without loss of generality, consider two classes, denoted by class $1$ and class $2$, such that $\theta_2>\theta_1$, and $\theta_1b<w$, $\theta_2b>w$. 
These conditions are general, and will be needed for the convenience of the canonical representation of the LP.
Rewrite~\eqref{1:lin_std} in matrix canonical representation:

\begin{align}
& \text{minimize } \no\\
&\;\;{\frac {h_{{1}} \left( w-\theta_{{2}}b \right) }{\theta_{1}-\theta_
{{2}}}}+{\frac {h_{{2}} \left( \theta_{{1}}b-w \right) }{\theta_{{1}}-
\theta_{{2}}}}+(h_{{3}}-h_1\frac{\theta_3-\theta_2}{\theta_1-\theta_2}-h_2\frac{\theta_1-\theta_3}{\theta_1-\theta_2})x_{{3}}+\cdots+(h_{I}-h_1\frac{\theta_I-\theta_2}{\theta_1-\theta_2}-h_1\frac{\theta_1-\theta_I}{\theta_1-\theta_2})x_{I}\no \\
& \;\;\text{   w.r.t } \nonumber\\
&\begin{pmatrix}
x_1 &   & {\frac { \left( \theta_{3}-\theta_{2} \right) }{\theta_{{1
}}-\theta_{{2}}}}x_{3}
 & {\frac { \left( \theta_{4}-\theta_{2} \right) }{\theta_{{1
 }}-\theta_{{2}}}}x_{4} & \cdots & {\frac { \left( \theta_{I}-\theta_{2} \right) }{\theta_{{1
 }}-\theta_{{2}}}}x_{I} \\
  & x_2  & {\frac { \left( \theta_{1}-\theta_{3} \right) }{\theta_{{1
}}-\theta_{{2}}}}x_{3}
 & {\frac { \left( \theta_{1}-\theta_{4} \right) }{\theta_{{1
 }}-\theta_{{2}}}}x_{4} & \cdots & {\frac { \left( \theta_{1}-\theta_{I} \right) }{\theta_{{1
 }}-\theta_{{2}}}}x_{I} \\\label{2:lin_std}
\end{pmatrix}
=
\begin{pmatrix}
\frac{b\theta_2-w}{\theta_2-\theta_1} \\
\frac{w-b\theta_1}{\theta_2-\theta_1}\\
\end{pmatrix}
\end{align}
In the case all the coefficients of the new objective function in the above display are  positive, by [Theorem 3.4.1,~\cite{thie_book}] the minimum for $\bar h$ is achieved at
\[
x=\{\frac{b\theta_2-w}{\theta_2-\theta_1},\frac{w-b\theta_1}{\theta_2-\theta_1},0,\cdots,0\},
\]
and is expressed by
\[\bar h=\frac {h_{1} \left( w-\theta_{2}b \right) }{\theta_{1}-\theta_
{{2}}}+{\frac {h_{2} \left( \theta_{1}b-w \right) }{\theta_{1}-
\theta_{2}}}\]
Alternatively, assume $(h_{{3}}-h_1\frac{\theta_3-\theta_2}{\theta_1-\theta_2}-h_2\frac{\theta_1-\theta_3}{\theta_1-\theta_2})$ is negative. Then, we act according to the simplex algorithm, as presented in~\cite{thie_book} and perform a pivot operation. This gives the following problem:
\begin{align}
& \text{minimize } \no\\
&\;\;{\frac {h_{1} \left( w-\theta_{3}b \right) }{\theta_1-\theta_
{3}}}+{\frac {h_{{2}} \left( \theta_{{1}}b-w \right) }{\theta_1-
\theta_{3}}}+(h_{{3}}-h_1\frac{\theta_2-\theta_3}{\theta_1-\theta_3}-h_2\frac{\theta_1-\theta_2}{\theta_1-\theta_3})x_2+\cdots+(h_{I}-h_1\frac{\theta_I-\theta_3}{\theta_1-\theta_3}-h_1\frac{\theta_1-\theta_I}{\theta_1-\theta_3})x_{I}\no \\
& \;\;\text{   w.r.t } \nonumber\\
&\begin{pmatrix}
x_1 &   & {\frac { \left( \theta_2-\theta_3 \right) }{\theta_1-\theta_3}}x_2
 & {\frac { \left( \theta_4-\theta_3 \right) }{\theta_1-\theta_{3}}}x_{4} & \cdots & {\frac { \left( \theta_{I}-\theta_{3} \right) }{\theta_1-\theta_3}}x_{I} \\
  & {\frac { \left( \theta_{1}-\theta_2 \right) }{\theta_1-\theta_3}}x_2 & x_3
 & {\frac { \left( \theta_{1}-\theta_{4} \right) }{{\theta_1}-\theta_3}}x_{4} & \cdots & {\frac { \left( \theta_{1}-\theta_{I} \right) }{\theta_1-\theta_3}}x_{I} \\\label{5:lin_std}
\end{pmatrix}
=
\begin{pmatrix}
\frac{b\theta_3-w}{\theta_3-\theta_1} \\
\frac{w-b\theta_1}{\theta_3-\theta_1}\\
\end{pmatrix}
\end{align}
One sees that the solution to this problem has a similar form. Hence, in the case all coefficients in the function to be minimized are positive, the minimum for $\bar h$ is achieved at this time
\[
x=\{\frac{b\theta_3-w}{\theta_3-\theta_1},0,\frac{w-b\theta_1}{\theta_3-\theta_1},0,\cdots,0\},
\]
and is expressed by
\[\bar h=\frac {h_{1} \left( w-\theta_{3}b \right) }{\theta_{1}-\theta_
{{3}}}+{\frac {h_{3} \left( \theta_{1}b-w \right) }{\theta_{1}-
\theta_{3}}}\]
See from~\cite{thie_book}, (section 3.8)  that there is a finite number of pivot steps which brings the canonical representation above to the form with objective function with non-negative coefficients. (There are additional options, such that the objective function is unbounded and/or degenerate cases which we rule out since they imply non-physical properties of the problem setting.)
Therefore, for any $w$, the minimal $\bar h(w)$ is achieved when \textit{only tasks belonging to at most two classes are present in the buffer}.
Note that the solution above is not necessarily unique, as more than one pair of $\theta_i$ may satisfy the conditions above. However, 
once initial workload is at $0$, the tasks of some class, denote it for a moment as $j$, are accumulated first. This class has the lowest $h_j/\theta_j$. That is, the first type of task to be accumulated in the buffer follows the well-known $c\mu$ rule, till it holds $w=b\theta_j$.
We derive next the rule how to find the two tasks which minimize the $\bar h$, for $w>b\theta_j$.
See that the maximal workload associated with tasks of (the cheapest) type $j$, is the lowest workload possible provided the buffer is full. 
Now assume that additional amount of workload worth of $\epsilon$ is added.
This workload addition is translated into reduction of number of tasks of type $j$ and addition of tasks of other types. 
Denote the number of tasks subtracted from type $j$ as $X$.
Then, the number of added tasks of all other types is $\sum_{i=1,i\neq j}^Ic_iX$, for non-negative $c_i$, such that $\sum_{i=1,i\neq j}^Ic_i=1$.
Total workload change is given by
\[
\epsilon=\sum_{i=1,i\neq j}^Ic_iX\theta_i-X\theta_j
\]
Then, the total number of tasks of type $j$ that were displaced (by saying displaced we mean served, and no additional tasks of that type were admitted) is
\[
X=\frac{\epsilon}{\sum_{i=1,i\neq j}^Ic_i\theta_i-\theta_j}
\]
We find the optimal cost associated with the workload addition worth of $\epsilon$, denote it as $P$.
\[
P=\sum_{i=1,i\neq j}^Ic_iXh_i-Xh_j=\frac{\epsilon(\sum_{i=1,i\neq j}^Ic_ih_i-h_j)}{\sum_{i=1,i\neq j}^Ic_i\theta_i-\theta_j}
\]
Alternatively, cost addition "per workload worth of $\epsilon$" is
\begin{equation}\label{m1}
P/\epsilon=\frac{\sum_{i=1,i\neq j}^Ic_ih_i-h_j}{\sum_{i=1,i\neq j}^Ic_i\theta_i-\theta_j}
\end{equation}
The following lemma is a straight-forward consequence of the theorem [3.4.1] mentioned above:
\begin{lemma}
The optimal combination of $c_i$ with respect to~\eqref{m1} is such that there exists one $c_k=1, k\neq j$ and $c_i=0$ for $i\neq k$.
 \end{lemma}
See that once the buffer is full with tasks of some type $j$, only tasks of the type with the \textit{lowest ratio} $\frac{h_k-h_j}{\theta_k-\theta_j}$ are admitted. That is, the additional cost accumulated due to the workload increase is minimal. Heuristically, as the workload continues to increase, this admission pattern continues till the buffer is full with tasks of type $k$, while all tasks of type $j$ have been displaced. Then, the tasks of the next type which comply to the similar condition are accumulated, instead of those of type $k$. The addition of the workload cannot go further if the free boundary is reached.
Note that the first class to be admitted (when the buffer is not full) is the one with the lowest ratio $\frac{h_k}{\theta_k}$.
We name this heuristically described procedure as {\it order of accumulation}. Figure~\ref{fig3} demonstrates the description above, in $4$ possible cases, for systems with $3$ and $2$ classes.

\begin{remark}\label{rem1}
For more consistency, one can define $\theta_0=0$ and $h_0=0$. Then, the expression $\frac{h_k-h_j}{\theta_k-\theta_j}$ will hold when the buffer is not full as well.
\end{remark}
We use these results to give a formal definition for the AO policy in the next section.

\subsubsection{Numerical example of Harrison-Taksar problem and its solution.}\label{sec:NumEx}

In what follows we present the numerical example.
The graphs below demonstrate the numerical solution of Harrison-Taksar problem for a shared buffer with maximal occupancy given by $b=125$ tasks, and other parameters given in Table~\ref{tab:tb1}.
We assumed $\hat\la_i=\hat\mu_i=0$ for all $i$ (so that $\bar m=0$), that
$\bar\sig^2=0.91$, and took the discount parameter $\al=10$.
The ordering of $r_i\mu_i$ is such that class $2$ is
the less expensive as far as rejections are concerned. Thus, this class is rejected at the free boundary. 

\begin{table}[h!]
\caption{\emph{Table of parameters of 3 sharing tasks}}
\begin{center}
\begin{tabular}{|c|c|c|c|c|c|c|}
\hline
Class & Holding cost & Rejection cost & $\mu$ & $\lambda$ &$r\mu$ \\
 \hline
  Class I  & 1390 & 962.5 & 1.80 & 0.60 & 1732.4  \\
  Class II  & 1050 & 700 & 2.20 & 0.73 & 1539.1\\
  Class III  & 733 & 875 & 2.80 & 0.93 & 2450.3\\
  \hline
\end{tabular}
 \end{center}
 \vspace{-0.3in}
\label{tab:tb1}
 \end{table}

The Bellman equation takes the form
\begin{equation}
  \label{25n}
  \begin{cases}
  \ds
 [2.19f''-10f+\bar h]\w f'\w[1539.19-f']=0,
 \qquad \text{in } (0,69.45),
 \\ \\
 f'(0)=0,\qquad
 f'(69.45)=1539.19.
 \end{cases}
\end{equation}
The function $\bar h$ is defined by
\[
\bar h(w)=\min\Big\{\sum_{i=1}^{3}h_i\xi_i:\xi\in\calX, \sum_{i=1}^{3}\theta_i\xi_i=w\Big\},
\qquad w\in[0,69.45],
\]
where $\calX=[0,125]$ and $\theta_i=\mu_i^{-1}$, which by numerical solution of the corresponding LP is translated into the following
\[
\bar{h}(w)\approx
\begin{cases}
  \ds
\frac{732.9\cdot w}{0.36}
 \qquad & 0\leq w\leq44.64,
 \\ \\
\ds
\frac{732.9(125\cdot0.45-w)}{0.45-0.36}+\frac{1050(w-125\cdot0.36)}{0.45-0.36}
  \qquad & 44.64<w\leq56.85,
  \\ \\
\ds
\frac{1050(125\cdot0.56-w)}{0.56-0.45}+\frac{1390(w-125\cdot0.45)}{0.56-0.45}
    \qquad & 56.85 < w \leq68.06.
\end{cases}
\]
Table~\ref{tab:tb2} demonstrates the solution to $\bar h$ and the selection of classes for each of the workload intervals. This selection takes part in setting the $\gamma$. Note that we used remark~\ref{rem1}.
\begin{table}[h!]
\begin{center}
\begin{tabular}{|c|c|c|c|c|}
\hline
$j$ & $\frac{h_1-h_j}{\theta_1-\theta_j}$ & $\frac{h_2-h_j}{\theta_2-\theta_j}$ & $\frac{h_3-h_j}{\theta_3-\theta_j}$ \\
 \hline
  0  & 2501.8 & 2308.7 & {\bf 2052.4}   \\
  3  & 3310.3 & {\bf 3245.6} & $\text{ }$ \\
  2  & {\bf 3.3730} & $\text{ }$ & $\text{ }$ \\
  \hline
\end{tabular}
 \end{center}
 \vspace{0in}
\caption{\sl\small\emph{Order of task accumulation. $j$ stands for the class which, in case the workload grows, fills up the buffer. The optimal selection, within each row, is in bold. As long as the buffer is not full, only class $I$ is admitted. Next, tasks of class $II$ are admitted. As the workload grows, tasks of class $I$ are displaced, while tasks of class $II$ fill up the buffer. Finally tasks of class $III$ ate accumulated, till free boundary is reached. We put blanks where the classes were already chosen for lower workload intervals.}}\label{tab:tb2}
 \end{table}
\begin{figure}[h!]\label{fig1}
\begin{center}
\includegraphics[width=40em]{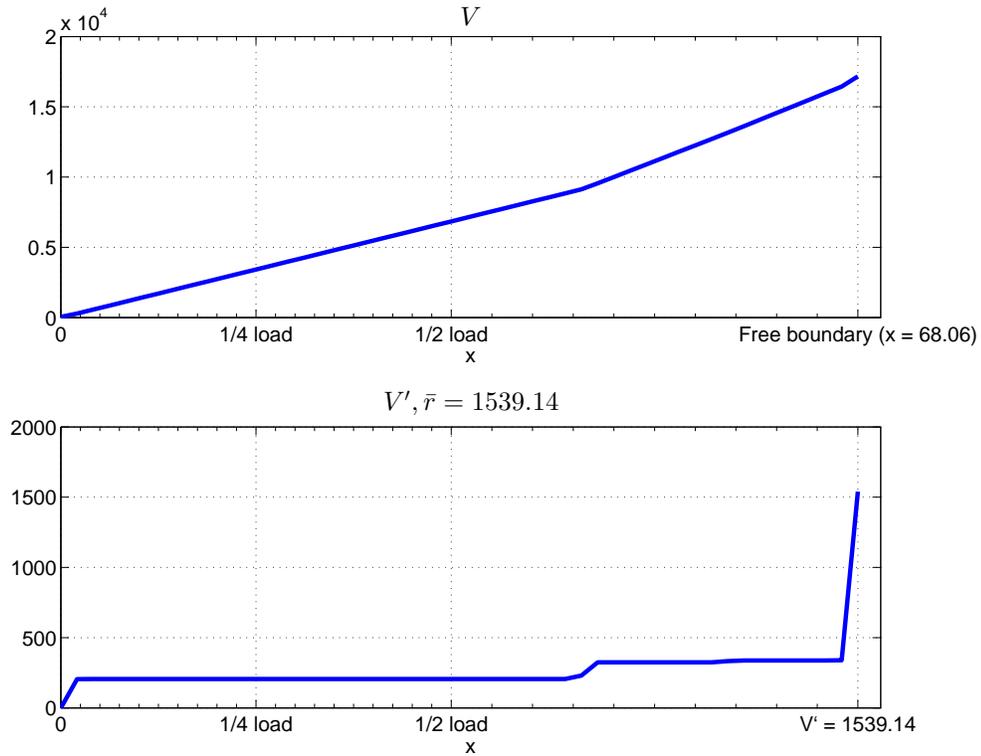}
\caption{\sl\small
$V$ and $V'$ within the free boundary. See that the maximal value of $V'$ is equal to $\bar r$ at free boundary.}
\end{center}
\end{figure}
\begin{figure}[h!]\label{fig2}
 \begin{center}
 \begin{align*}
 \includegraphics[width=30em]{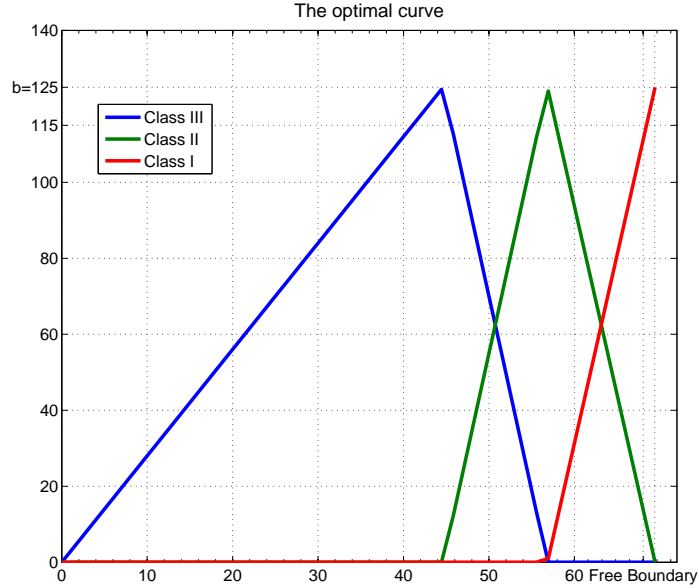}
 \end{align*}
 \caption{\sl\small
Classes which form the optimal curve $\gamma$. X-axis stands for the workload, while Y-axis shows the number of tasks of each class present in the buffer at this workload. The optimal queue-lengths at free boundary are $\{0,27.5595,97.4405\}$ (dashed vertical line).}
 \end{center}
 \end{figure}

The optimal curve $\gamma$ is given by the following
\begin{align*}
\gamma(w)\approx
\begin{cases}
  \ds
\frac{w}{44.64}125\, [0, 0, 1]
 \qquad & 0\leq w\leq44.64,
 \\ \\
 \ds
\frac{125*0.45-w}{0.45-0.36}\, [0,0, 1]+\frac{w-125*0.36}{0.45-0.36}\, [0, 1, 0]
  \qquad & 44.64<w\leq56.85,
  \\ \\
  \ds
\frac{125*0.56-w}{0.56-0.45}\, [1, 0, 0]+\frac{w-125*0.45}{0.56-0.45}\, [0, 1, 0]
    \qquad & 56.85 < w \leq68.06.
\end{cases}
\end{align*}

\section{A nearly optimal policy}\label{sec4}
We introduce the curve $\gamma$ for the triangular domain, given by~\eqref{10}.
The parameter $\bx$ associated with the RBCP is given by $\theta_{m}b$, where $m=\argmax_i{\theta_i}$. This corresponds to the buffer being full with most processor-consuming tasks. The tasks of this class contribute the maximal per-task workload.
In what follows, we label the classes according to their \textit{order of accumulation}, which was heuristically described in previous section. Again, assume that the workload grows from $0$ to $\bx$. This ordering will form the \textit{accumulation priorities} $p(j)$, and
it will be used to formulate the mapping from $w$ to $\gamma(w)$.
The numbering $j\in\{1,\cdots,I\}$, refers to the workload points $\hat w_j$, where the buffer is solely full with tasks of some type $i=p(j)$, such that $\hat w_j<\hat w_{j+1}$. (Note that in sequel we redefine this ordering in order to obtain $i=p(i)$.)
$\hat w_0=0$, stands for the zero workload.
$\hat w_1$ is defined as $w_1$ in the previous subsection, i.e the minimal load then the buffer is full. Namely,
\[
\hat w_1=\theta_nb, \;\;\;  n=\argmin_i{h_i\mu_i}, \; i\in\{1,\cdots,I\}
\]
The calculation of  $p(j)>1$ is performed recursively.
\[
p(j)=\argmin_i\frac{h_i-h_{p(j-1)}}{\theta_i-\theta_{p(j-1)}} \; : h_i>h_{p(j-1)}, \theta_i>\theta_{p(j-1)}
\]
Observe that the restriction $h_i>h_{p(j-1)}, \theta_i>\theta_{p(j-1)}$ means that no class can be chosen twice. (For otherwise that $i$ would be chosen for the lower $p(j)$). In addition, the condition on $\theta_i$ assures that the workload added by each task of class $p(j)$ is higher than that of task of class $p(j-1)$. See that $|p(j)|\leq I$.
Note that some classes can never be optimal, (for example, those with comparatively high $h_i$ and low $\theta_i$) and thus be never accumulated. However, the tasks of class with maximal $\theta_i$ are always assigned to the last $j$.
Note that by remark~\ref{rem1} one can define $h_{p(0)}=0$ and $\theta_{p(0)}=0$ to have the order of accumulation started with an empty buffer.

Define two groups of tasks:
\begin{align*}
&\calE=\{i| \exists j:p(j)=i\} \\
&\calD=\{i| \nexists j:p(j)=i\}
\end{align*}
Denote $p_m=|\calE|$. The group $\calD$ constitutes the task types which are always at low accumulation priority. See that $\calD$ can be an empty set.
Next, for simplicity of notation, we {\it reorder} the class indexing such that
$i=p(i),\; \text{if } i\in\calE$ and $i=\{p_m+1,p_m+2,\cdots,I\},\; \text{if } i\in\calD\}$. That is, the tasks are rearranged in the order of their accumulation in the buffer (i.e. the order of $i$ is equivalent to the order of $j$ before the reordering). Denote $J=|\calE|$.
The workload intervals are denoted by $\omega_j$.
\begin{equation}\label{91}
\omega_j=[\hat w_{j-1},\hat w_j),\; j>1
\end{equation}
Given $w\in[0,\bx]$, $(j,\xi_l,\xi_h)=(j,\xi_l,\xi_h)(w)$ are determined by $w\in\omega_j$ and we use the solution to~\eqref{2:lin_std} to set
\begin{equation}\label{3:lin_std}
\xi_h(w)=\frac{w-b\theta_{j-1}}{\theta_j-\theta_{j-1}}, \; \;
\xi_l(w)=b-\xi_h(w),
\end{equation}
in case $j>1$, and
\begin{equation}\label{4:lin_std}
\xi_l(w)=0, \; \;
\xi_h(w)=w/\theta_1,
\end{equation}
in case $j=1$.
With this notation, $\gamma$ is given by
\[
\gamma(w)=\xi_h\theta_j+\xi_l\theta_{j-1}.
\]
Clearly, the curve, which is expressed by $\gamma$ defined above, can lie along the boundary $\pl\calX$ of $\calX$, where it holds $\pl^+\calX:=\{x\in\calX:\sum_ix_i=b \}$, i.e. the buffer is full.
Note that $\gamma$ is the solution to the BCP which, in addition, comes in concert with the free boundary solution found from the RBCP, according to which only rejections at the workload level $\bx^*$ are allowed. Thus, these two properties of the solution can be seen as contradicting, when treating the QCP.
Hence, we propose a policy which approximates $\gamma$ by introducing an alternative curve, which is closed to $\gamma$ and is bounded away from the buffer limit boundary at the same time.
Let $\eps\in(0,b)$ be given.
Let $a=b-\eps$, and
$a^*:=\bx^*\w(\theta_Ja)<\bx=\theta_Jb$. In the case $\eps$ is small, we have
then $a^*=\bx^*$ (unless $\bx^*=\bx$).
Note that the buffer can be full for various values of $\xi_l,\xi_h$.
Therefore, we define additional margins that can vary according to the workload $w$.
Let $\eps_l=\eps_l(w)$ and $\eps_h=\eps_h(w)$. For each $w$ we set
$\eps_l$ and $\eps_h$ as follows
\begin{align}\label{2m}
&\eps_l=\begin{cases}
\eps_l=min\{\eps/2,\xi_l\}, & \text{if } \xi_l<\xi_h,\\
\eps_l=\eps/2, & \text{if } \xi_l\geq\xi_h,\;\xi_h>\eps/2 \\
\eps_l=\eps-\xi_h, & \text{if } \xi_l\geq\xi_h,\;\xi_h\leq\eps/2
\end{cases} \nonumber \\
&\eps_h=\eps-\eps_l
\end{align}
Let $\chi_l=\xi_l-\eps_l$ and $\chi_h=\xi_h-\eps_h$.
The approximation $\gamma^a:[0,\bx]\to\calX$ of $\gamma$ is as follows.
For $w\in[0,\theta_Ja)$, the variables $j=j(w)$ and $\chi_l=\chi_l(w),\chi_h=\chi_h(w)$ are determined via
\begin{equation}\label{61}
w=\theta_{j-1}\chi_l+\theta_j\chi_h, \qquad j\in\{1,2,\ldots,J\},\;\chi_l+\chi_h\in[0,a),
\end{equation}
and
\begin{equation}\label{62}
\gamma^a(w)=\chi_le^{(j-1)}+\chi_he^{(j)}.
\end{equation}
Note that the triplet $(j,\chi_l,\chi_h)$ is unique. We will refer to it as the {\it representation $(j,\chi_l,\chi_h)$ of $w$ via \eqref{61}}.
Denote $w_\chi=\theta_l\chi_l+\theta_h\chi_h$ and $w_\xi=\theta_l\xi_l+\theta_h\xi_h$, where $\theta_l$ and $\theta_h$ refer to the classes which are associated with $\xi_l$ and $\xi_h$, correspondingly. We need  the function $\gamma^a$ to be continuous on $[w_\chi,w_\xi]$ and satisfy the relation
$\theta\cdot\gamma^a(w)=w$. Hence, we define the linear interpolation between the points $w_\chi$
and $w_\xi$ as follows:
\begin{equation}\label{62int}
\gamma^a(w)=a+\frac{w-w_\chi}{w_\xi-w_\chi}(b-a),
\qquad w\in[w_\xi,w_\chi].
\end{equation}



We next specify the policy.
$Z^n(t)$ defines the rejection policy, while $B^n(t)$ describes the server allocation policy,
as a function of $X^n(t)$. In what follows the policy is set considering the index correction after the reordering above.

\subsubsection*{Rejection policy:}
The multidimensional rejection process $Z$ has only one nonzero component,
namely the $i^*$-th component, which increases only when $\bar X\ge\bx^*$.
This structure is translated to the queueing model. Hence, we use it to
construct the rejection component of the asymptotically optimal policy.
The latter occurs when $\theta\cdot \hat X^n\ge a^*$.
The forced rejections only occur in order to keep the
buffer size constraint.

\subsubsection*{Service policy:}
\begin{figure}[h!]
\begin{center}
\begin{align*}
\includegraphics[width=20em]{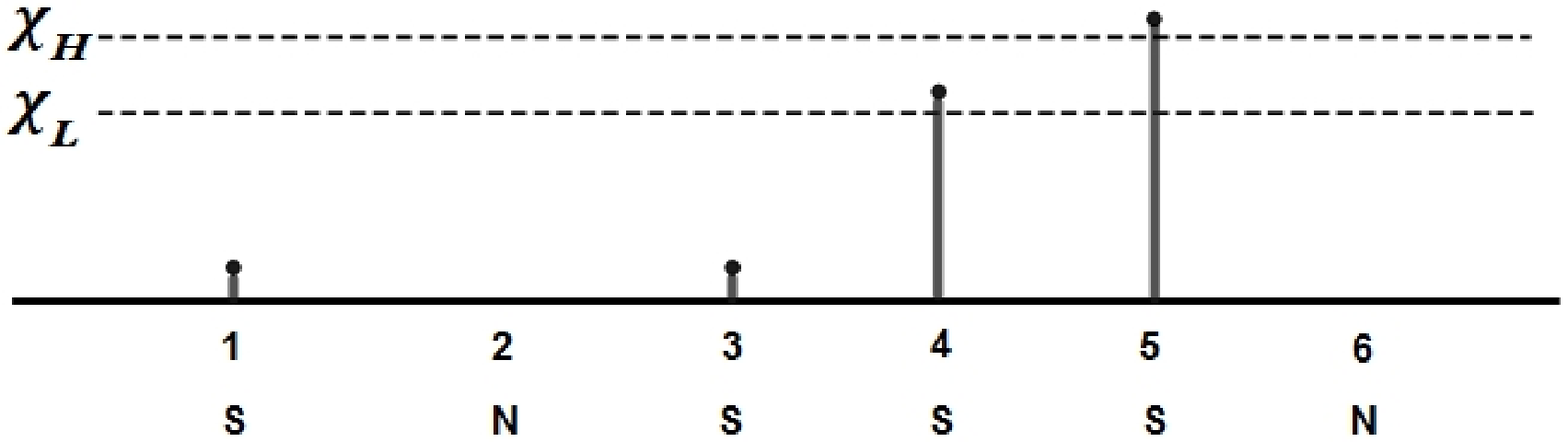}
\hspace{2em}
&
\hspace{2em}
\includegraphics[width=20em]{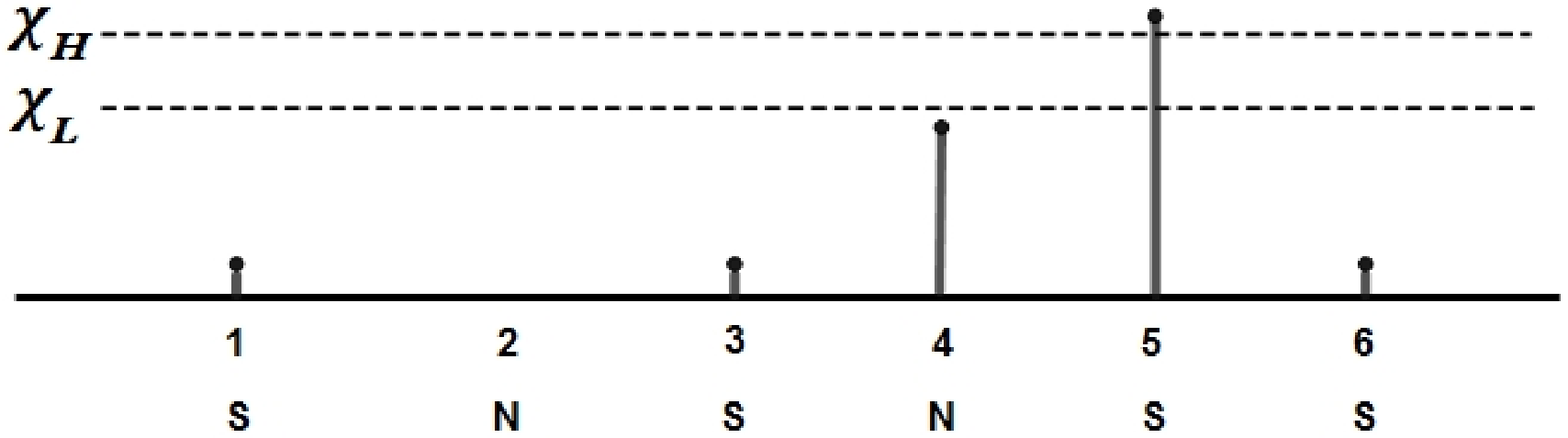}
\end{align*}

\caption{\sl\small
Schematic example of service allocation.
The figures depict possible states $\hat X^n(t)=x$. Classes denoted by $S$ are being served, classes denoted by $N$ are not being served.
}\label{fig5}
\end{center}
\end{figure}

For each $x\in\calX$ define the low priority:
\[
\calL(x)=\{j | x_j<\chi_h, \;\; j-1 | x_{j-1}<\chi_l\}
\]
That is, the tasks of type $j$ and $j-1$, where $j=j(w)$ have always low priority as long as their quantity is below $\chi_l$ and $\chi_h$, correspondingly.
The complement set defines the high priority classes:
\[
\calH(x):=\calI\setminus\{\calL(x)\}.
\]
Denote $\calH^+(x)=\{i\in \calH(x):x_i>0\}$ and $\calL^+(x)=\{i\in \calL(x):x_i>0\}$.
The policy is to allocate service to all classes within $\calH^+(x)$ equal to a fraction proportional to the corresponding traffic intensities. 
Classes within $\calL(x)$ receive no service, with exception of the moments then $\calH^+(x)$ is empty. 
\begin{equation}\label{58}
\rho^L_i(x)=\begin{cases}
0, & \text{if } 1_{(x_i=0)\cup(\calH^+(x)\neq\emptyset)}=1 , \\
\ds\frac{\rho_i1_{\{i\in\calL^+(x)\}}}{\sum_{k\in\calL^+(x)}\rho_k}, &
\text{if } \calH^+(x)=\emptyset\\
\end{cases} \qquad
\rho^H_i(x)=\begin{cases}
0, & \text{if } x_i=0,\\
\ds\frac{\rho_i1_{\{i\in\calH^+(x)\}}}{\sum_{k\in\calH^+(x)}\rho_k}, &
\text{if } \calH^+(x)\ne\emptyset,\\
\end{cases}
\end{equation}
Define $\rho'_i(x)=\rho^L_i(x)\cdot1_{i\in\calL}+\rho^H_i(x)\cdot1_{i\in\calH}$.
Then for each $t$,
\begin{equation}
  \label{32}
  B^n(t)=\rho'(\hat X^n(t)).
\end{equation}
Note that when $\calH^+(x)\ne\emptyset$,
\begin{equation}
  \label{31}
  \rho'_i(x)>\rho_i\quad \text{for all } i\in \calH^+(x).
\end{equation}
That is, all classes which receive service are allocated
a fraction of effort strictly greater than their traffic intensity.
See that $\sum_iB^n_i=1$ whenever $\hat X^n$ is nonzero. Hence, the proposed policy is work conserving.
Figure~\ref{fig5} demonstrates two cases of service policy for $6$ classes. The reader may also consider to go back to Table~\ref{tab:tb2} to review the example of the order of accumulation.

\begin{remark}
  \label{rem4}
  The tasks of type $i>J$ (if exist) have always high priority once arrive to the system. This is because these tasks are never optimal to have them accumulated in the buffer, according to the solution of $\bar h$. However, one of these task types may be chosen to be rejected at free boundary, according to the $\bar r$.
\end{remark}
\begin{remark}
Note, that the rejection policy resembles that of the rectangular case, because the calculation of $\bar r$ follows the same rule. The class $i^*$ is the only class which is subject to the non-forced rejection.
\end{remark}

\section{State space collapse.}\label{sec:ssc}

In this section, we prove the main theorem which states the upper bound for the value function under proposed policy. The theorem is written as follows.
\begin{theorem}
  \label{th2}
  For each $\eps>0$ and $n$, denote the policy constructed above by $U^n(\eps)$.

  Then,

   $\limsup_{n\to\iy}J^n(U^n(\eps))\le V(x_0)+\al(\eps)$, where $\al(\eps)\to0$ as $\eps\to0$.
\end{theorem}

\begin{remark}
The combination of Theorem~\ref{th2} above and Theorem 3.1 in~\cite{Atar-Shif} (which states the general lower bound) provides the AO of the policy $U^n(\eps)$.
\end{remark}

\textit{Proof.}

For fixed $\eps$ in $U^n(\eps)$, write $U^n=(Z^n,B^n)$ and
denote by $\tau^n$ the time of the first forced rejection. We perform most of the analysis on the processes before $\tau^n$ is reached. We show that the cost of QCP, under proposed policy, weakly converges to the value function associated with BCP solution as $n\to\iy$.
We divide the proof into two major steps. First step shows that the workload process
$\theta^n\cdot\hat X^n$ converges to RBM. This results in that only rejections from class $i^*$ occur, and only when $\theta^n\cdot\hat X^n\approx a^*$.
The second step, shows that $\hat X^n$ lies close to the minimizing curve at all times. Hence, the running cost is locally minimized.
This establishes that in any finite time, $\tau^n$ is not reached.

We prove for the case where the system starts with initial condition
close to the minimizing curve, namely
\begin{equation}\label{49}
\hat X^n(0)-\gamma^a(\theta\cdot\hat X^n(0))\to0 \text{ as $n\to\iy$,
and }\theta^n\cdot\hat X^n(0)\in[0,a^*] \text{ for all $n$ large.}
\end{equation}
The assumption on initial condition can be relaxed in a straightforward manner. Shortly, given a general initial condition, there will be a jump towards a position located on the curve. We skip the technical details which can be found in~\cite{Atar-Shif}.

Denote some $\eps'$, a distance from the minimizing curve $\gamma^a(\theta\cdot x)$, small enough to assure that forced rejections within that distance do not occur.
Observe that $\sig^n\le\tau^n$, where
\[
\sig^n=\hat\zeta^n\w\zeta^n,
\]
\[
\hat\zeta^n=\inf\{t:X^{\#,n}\ge a^*+\eps'\},\qquad
\zeta^n=\inf\{t:\max_{i\le I}|\Del^n_i(t)|\ge\eps'\}.
\]
In this sense, $\hat\zeta^n$ refers to the time of violating of the free boundary of the workload,
while $\zeta^n$ refers to the time when distance of number of tasks of any tasks-type from the
optimality curve overcomes $\eps'$.

Now multiply equation \eqref{17} by the vector $\theta^n=(1/\mu^n_i)_{i\in\calI}$ and
denote
\begin{equation}
  \label{33}
  W^{\#,n}=\theta^n\cdot\hat W^n,\quad
  X^{\#,n}=\theta^n\cdot\hat X^n,\quad
  Y^{\#,n}=\theta^n\cdot\hat Y^n,\quad
  Z^{\#,n}=\theta^n\cdot\hat Z^n.
\end{equation}
We have
\begin{equation}
  \label{34}
  X^{\#,n}=X^{\#,n}(0)+W^{\#,n}+Y^{\#,n}-Z^{\#,n}.
\end{equation}
Let $W^{\circ,n}:=W^{\#,n}(\cdot\w\tau^n)$ denote the process $W^{\#,n}$ when
stopped at the time $\tau^n$. Define similarly $X^{\circ,n}$, $Y^{\circ,n}$
and $Z^{\circ,n}$.

The following lemma states that the free boundary is not  violated (that is, $\hat\zeta^n$ is not reached).
\begin{lemma}\label{lem4}
any subsequential limit $(\tilde W,\tilde X,\tilde Y,\tilde Z)$ of the  sequence
$(W^{\circ,n},X^{\circ,n},Y^{\circ,n},Z^{\circ,n})$ is
$C$-tight, and satisfies a.s.,
\begin{equation}\label{43}
(\tilde X,\tilde Y,\tilde Z)=\Gam_{[0,a^*]}[\bar x_0+\tilde W].
\end{equation}
\end{lemma}
The detailed proof is given in~\cite{Atar-Shif}. (See Theorem $4.1$, {\it Step 1}).

Hence, we straightly proceed to prove our main result, which constitutes the proof of the state-space collapse. 
Our objective is to show that the multidimensional process $\hat X^n$
lies close to the minimizing curve. That is, as $n\to\iy$,
\begin{equation}
  \label{35}
  \Del^n(t):=\hat X^n(t)-\gamma^a(X^{\#,n}(t))\To0,
\end{equation}
uniformly on compacts.

It suffices to show that $\PP(\sig^n<T)\to0$, for any small $\eps'>0$ and any $T$.
Fix $\eps'$ and $T$. Thanks to the fact that $\sig^n\le\tau^n$,
{\small
\begin{equation}\label{60}
\PP(\sig^n<T)\le\PP(\hat\zeta^n\w\zeta^n\le T\w\tau^n)
\le\PP(\hat\zeta^n\le T\w\tau^n)+\PP(\zeta^n\le T\w\tau^n).
\end{equation}
}
From lemma~\ref{lem4} it follows that
$\PP(\hat\zeta^n\le T\w\tau^n)\to0$ as $n\to\iy$.
It therefore suffices to prove the following lemma.

\begin{lemma}\label{lem3}
$\PP(\zeta^n\le T\w\tau^n)\to0$ as $n\to\iy$.
\end{lemma}

\textit{Proof.}

On $\zeta^n\le T\w\tau^n$ let
$x^n:=X^{\#,n}(\zeta^n)=X^{\circ,n}(\zeta^n)$ and let $j=j^n$ and $\chi_l^n,\chi_h^n$ be the corresponding
components from the representation $(j,\chi_l,\chi_h)$ of $x^n$ (with $w=x^n$).

The proof strategy is to define a covering of the workload domain $[0,\bx]$ by intervals. Then, we assume $w$ being present in one of these intervals and prove the lemma for all possible cases.
More precisely, we show that it is sufficient to distinguish between four different types of intervals in order to treat all possibilities.

To that end, fix a positive integer $K=K(\eps')=[c_0/\eps']$, where $c_0$ is a constant depending only
on $\theta$. ( we treat this value at a later stage of the proof.)
Define
$K-1$ intervals $\X_k=\bB(k\eps_1,\eps_1)$, $k=1,2,\ldots,K-1$, where $\bB(x,d)$ denotes a closed interval
$[x-d,x+d]$ and $\eps_1=\bx/K$. Let $\tilde\X_k=\bB(k\eps_1,2\eps_1)$.

We use the characterization of $C$-tightness
as in Proposition VI.3.26 of \cite{jac-shi} and apply it to $X^{\circ,n}$. Given $\del>0$ there exists
$\del'=\del'(\del,T,\eps_1)>0$, such that for all sufficiently large $n$,
\begin{equation}
  \label{39}
  |X^{\circ,n}(s)-X^{\circ,n}(t)|\le\eps_1 \text{ for all } s,t\in[0,T], |s-t|\le\del',
  \text{ with probability at least } 1-\del.
\end{equation}
Fix such $\del$ and $\del'$. Denote by $\bT^n$ the interval $[(\zeta^n-\del'\vee 0),\zeta^n]$. Define the following event,
\begin{equation}\label{5m}
\Om^{n,k}=\{\zeta^n\le T\w\tau^n,x^n\in\X_k,X^{\#,n}(t)\in\tilde\X_k\text{ for all }
t\in\bT^n\}.
\end{equation}
By simple probabilistic manipulations, using that $X^{\#,n}=X^{\circ,n}$ on $[0,\tau^n]$, it can be shown that for all large $n$,
\begin{equation}\label{63}
\PP(\zeta^n\le T\w\tau^n)\le\del+\sum_k\PP(\Om^{n,k}),
\end{equation}
We fix the index $k$, which is associated with the $k$-th interval in the workload covering, and analyze $\Om^{n,k}$. We aim to show that $\PP(\Om^{n,k})\to0$ as $n\to\iy$, for each $k$.
The value assigned by the policy to $B^n$ (see \eqref{32}) remains fixed as
$\hat X^n$ varies within any of the intervals $(\omega_j)$ defined in~\eqref{91}. However, this is not necessary the case. There are several cases, which we separately define and treat as it follows below. 

\subsubsection*{\textbf{(I)}} $\tilde\X_k\subset(0,a^*)$ and for all $j$, $\hat w_j\notin\tilde\X_k$. That is, $\hat X^n$ remains
in the same interval during the time window $\bT^n$. Hence, the region $\omega_j$ is constant. 
To prove in this case, see that all points $x$ in
$\tilde\X_k$ are translated to the same $j$ in the representation $(j,\chi_l,\chi_h)$ of $x$, as in
\eqref{61}.
Note that $j=j(k)$ depends on the interval $k$ only, and does not vary with $n$.
Also, $j=j^n$ under $\Om^{n,k}$.
\AddedTh{
We split the proof into two separate cases corresponding to the two groups of classes - $i\notin\{j-1,j\}$,
$i\in\{j-1,j\}$. We also separately treat the case where $j=1$, i.e. the buffer is not full. We start with the first group.}
Fix $i\notin\{j-1,j\}$. 
We estimate the probability that, on $\Om^{n,k}$, $\zeta^n\le T\w\tau^n$ occurs by
having $\Del^n(\zeta^n)\ge\eps'$.
More precisely, note that $\gamma^a_i(x^n)=0$ (this follows from the policy definition in~\eqref{58}, i.e. because $i\notin\{j-1,j\}$).
Then we will show that
\begin{equation}\label{41}
\text{for every } \eps''\in(0,\eps'),\quad
\PP(\Om^{n,k}\cap\{\hat X^n_i(\zeta^n)>\eps''\})\to0\quad \text{ as } n\to\iy.
\end{equation}
Note that $\gamma^a$ is continuous and that $\Del^n(0)\to0$ as $n\to\iy$,
by \eqref{49}. Using the fact that the jumps
of $\hat X^n$ are of size $n^{-1/2}$, on the event indicated in \eqref{41} there must exist
$\eta^n\in[0,\zeta^n]$ with the properties that
\begin{equation}\label{55}
\hat X^n_i(\eta^n)<\eps''/2,\qquad X^n_i(t)>0 \text{ for all } t\in[\eta^n,\zeta^n].
\end{equation}
Define $\hat\eta^n=\eta^n\vee(\zeta^n-\del')$. We examine now
$\hat X^n_i[\zeta^n]-\hat X^n_i[\hat\eta^n]=\hat X^n_i[\hat\eta^n,\zeta^n]$, aiming to bound it using the modulus of continuity. Since the probability of the latter to be positive should go to zero, the probability that the difference $\hat X^n_i[\hat\eta^n,\zeta^n]$ is positive will go to zero as well. See that due to the position of $\hat X^n(t)$ in interval $\tilde\X_k\subset(0,a^*)$ (by definition of case (\textit{I\textbf{}})) there are no rejection at free boundary at all times in time interval $[\hat\eta^n,\zeta^n]$. Hence, $\hat Z^n_{i^*}[\hat\eta^n,\zeta^n]=0$. Since the buffer is full, we must show that the probability of the forced rejections goes to zero as $n\to\iy$. for all $i$. Assume that jobs of classes $j-1$ and $j$ are bounded away from $\xi_l$ and $\xi_h$ respectively. Hence, the arrivals of class $i$ are admitted and $Z^n_{i}[\hat\eta^n,\zeta^n]=0$, with high probability.
Observe that on the event in~\eqref{55}, during the time interval $[\eta^n,\zeta^n]$,
$i$ is always a member of $\calH(\hat X^n)$.
Recall (from definition in~\eqref{2}) $T^n_i(t)=\int_0^t B^n_i(s)ds$, while
by \eqref{32}--\eqref{31}, $B^n_i(t)=\rho'_i(\hat X^n(t))>\rho_i+c$, for some constant $c>0$. 
Substitute in \eqref{18} and make time derivative 
\be\label{92}
\frac{d}{dt}\hat Y^n_i(t)\leq\frac{d}{dt}(\frac{\mu^n_i}{\sqrt n}(\rho_it-\int_0^t (\rho_i+c)ds))=-\frac{\mu^n_i}{\sqrt n}c
\ee
\AddedTh{Note that definition of $\hat\eta^n$ assures that the equations which follow next will be valid on $\Omega^{n,k}$, i.e. within the time interval $\bT^n$. That is, by~\eqref{63} and~\eqref{4m} we have that the time interval $[\hat\eta^n,\zeta^n]$ is bounded by $\delta'$, while $\hat X^n_i[\hat\eta^n,\zeta^n]$ on it is bounded by $\eps_1$.}
Using these facts in \eqref{17} and substituting $(\zeta^n-\hat\eta^n)$ for the time interval in~\eqref{92}, we have
\begin{equation}\label{4m}
\hat X^n_i[\hat\eta^n,\zeta^n]=\hat W^n_i[\hat\eta^n,\zeta^n]
-c\frac{\mu^n_i}{\sqrt n}(\zeta^n-\hat\eta^n).
\end{equation}
Fix $r_n$, a positive sequence, such that $r^n\to0$ and $r^n\sqrt n\to\iy$. We use this sequence to indicate the growth of $\zeta^n-\hat\eta^n$ with $n$, referring to two different cases.
The $C$-tightness of $\hat W^n_i$ provides a bound by means of modulus of continuity; that is, $w_T(\hat W^n_i;r_n)\ge\hat W^n_i[\eta^n,\zeta^n]$.

In the first case, we assume $\zeta^n-\eta^n<r_n$ and $n$ is sufficiently large, such that $\hat\eta^n=\eta^n$. Moreover, $c\frac{\mu^n_i}{\sqrt n}(\zeta^n-\hat\eta^n)$ is finite and positive.
Thus by definition of $\eta^n$, and by the assumption that jump of size $\eps''/2$ happened, we have $\hat X^n_i[\hat\eta^n,\zeta^n]\ge\eps''/2$. Consequently,
\[
w_T(\hat W^n_i;r_n)\ge\hat W^n_i[\eta^n,\zeta^n]\ge\eps''/2
\]
must hold. However, the probability of this event goes to zero as $n\to\iy$. 

In the second case we assume that $\zeta^n-\eta^n\ge r_n$. Hence, by \eqref{4m} we omit $\eps''/2$ and change sides,
\[
2\|\hat W^n_i\|_T\ge\hat W^n_i[\hat\eta^n,\zeta^n]\ge c\frac{\mu^n_i}{\sqrt n}r_n
\ge cr_n\sqrt n,
\]
for some constant $c>0$. Observe that since $r_n>0$, the probability of the event above goes to zero as well.
Summarizing, the probability in \eqref{41} is bounded by
\begin{equation}
\label{53}
\PP(w_T(\hat W^n_i;r_n)\ge\eps''/2)+\PP(2\|\hat W^n_i\|_T\ge cr_n\sqrt n),
\end{equation}
which converges to zero as $n\to\iy$, by $C$-tightness of $\hat W^n$. This proves \eqref{41}. Consequently, the classes which do not belong to the $\{j-1,j\}$ stay at zero with high probability. Moreover, forced rejections of these classes happen with low probability.

Note that in the case $i\notin\{j-1,j\}$ we demonstrated that $X^{n}(t)$ lies near zero. Hence, we could assume that $\gamma^a_j(X^{\circ,n}=0$. 
In order to prove for the case $j-1$ and $j$ we aim to show, 
\begin{equation}\label{52}
\text{for every } \eps''\in(0,\eps'),\quad
\PP(\Om^{n,k}\cap\{\Del^n_j(\zeta^n)>\eps''\})\to0
\quad \text{ as } n\to\iy.
\end{equation}
That is, classes $j-1$ and $j$ are bounded away from $\xi_l$ and $\xi_h$ by predefined constants $\varepsilon_l$ and $\varepsilon_h$, respectively.
Therefore, we will show that the distance from the curve, $\Del^n_j[\hat\eta^n,\zeta^n]$, goes to zero with high probability.
For simplicity of notation, define $C^n(t)=\gamma^a_j(X^{\circ,n}(t))$. 
Then $\Del^n_j=\hat X^n_j-C^n$.
Recall that by policy definition for $i=j$,
\begin{align}
  \text{$j\in\calH(\hat X^n(t))$ 
  whenever
  $\hat X^n_j(\eta^n)\geq\chi_h$.} \label{51}\\
  \text{$j\in\calL(\hat X^n(t))$ 
    whenever
    $\hat X^n_j(\eta^n)<\chi_h$.} \no 
\end{align}
We refer to the first option first. Similarly to \eqref{55}, there exists $\eta^n\le\zeta^n$ such that
\begin{equation}\label{55a}
\hat X^n_j(\eta^n)<\chi_h+\eps''/2,\qquad \hat X^n_j(t)>\chi_h \text{ for all } t\in[\eta^n,\zeta^n].
\end{equation}

Again, we distinguish between two cases: $\zeta^n-\eta^n<r_n$ and
$\zeta^n-\eta^n\ge r_n$. We have now two terms of modulus of continuity, referring to the processes $W^n_i$ and $C^n$. Recall that we by definition, $\gamma^a_j$ is continuous (\eqref{62}--\eqref{62int}) and by Lemma~\ref{lem4} $X^{\circ,n}(t)$ is $C$-tight. Hence, $C^n$ is $C$-tight. Therefore, proving similarly as for $i\notin\{j-1,j\}$, the probability in \eqref{52} is bounded by
\begin{equation}
\label{54}
\PP(w_T(\hat W^n_i;r_n)+w_T(C^n;r_n)\ge\eps''/2)
+\PP(2\|\hat W^n_i\|_T+2\|C^n\|_T\ge cr_n\sqrt n),
\end{equation}
Note that in the second option of~\eqref{51} class $i$ is not served and thus jumps immediately to the point $\chi_h$ with high probability, as $n\to\iy$. 
This accomplishes the proof for $i=j$. The proof for $i=j-1$ is similar, with the only difference that $\chi_h$ is substituted by $\chi_l$.
We now separately treat the case $j=1$. For this case $j-1=0$ and $\chi_l=0$, which is treated as for the task classes $i\notin\{j-1,j\}$, and~\eqref{41} holds. The proof for the $j$ itself is similar to the case where $j>1$ and~\eqref{52} holds. Thus,
\begin{equation}\label{52a}
\text{for every } \eps''\in(0,\eps'),\quad
\PP(\Om^{n,k}\cap\{\Del^n_{j-1}(\zeta^n)>\eps''\})\to0
\quad \text{ as } n\to\iy.
\end{equation}
We can now show that $\PP(\Om^{n,k})\to0$ as $n\to\iy$. 
See that by definition of $\gamma$, the minimizing cost, and by the convergence established for the workload process
$\theta\cdot\gamma^a(\theta\cdot x)=\theta\cdot x$ for all $x\in\calX$ and $\theta^n\to\theta$. Combining this with~\eqref{41}~\eqref{52},~\eqref{52a}, we bound $|\Del^n_i(\zeta^n)|$ as follows
\begin{equation}
\label{57}
\PP(\Om^{n,k}\cap\{\max_{i\le I}|\Del^n_i(\zeta^n)|>\eps''\})\to0,
\end{equation}
Since $\eps''$ is arbitrarily small, it follows from the definition of $\zeta^n$
that $\PP(\Om^{n,k})\to0$ as $n\to\iy$.
\subsubsection*{(\textbf{II})} We consider the case
$\tilde\X_k\subset(0,a^*)$
but $\hat w_j\in\tilde\X_k$ for some $j\in\{1,2,\ldots,J\}$. That is, the buffer is \textit{mostly} filled with tasks of class $j$. 
Let $(j^n(t),\xi^n(t))$ denote the representation \eqref{61} for $X^{\#,n}(t)$.
In the case $j^n>2$, in the
time window $\bT^n$, $j^n,(j-1)^n$ vary in the range of $j-1$, $j$ and $j+1$. More precisely, we have to analyze the fluctuation between $\{j-1,j\}$ and $\{j,j+1\}$. The case $j^n=2$, stands for fluctuation between $1$, (then buffer is not nearly full) and $\{1,2\}$.  In case $j^n=1$ we have $\hat w_0\in\tilde\X_k$ which is separately treated in case \textbf{(\emph{III})}. 

We first treat the case where $2\leq j<J$.
In this case, the buffer is mostly filled with tasks of class $j$. See by~\eqref{3:lin_std} that if $w>\hat w_j$, the tasks of class $j$ are associated with $\chi_l$. Define
\[
\chi_h^{(1)}(w)=\frac{w-a\theta_{j}}{\theta_{j+1}-\theta_{j}}, \; \;
\chi_l^{(1)}(w)=a-\chi_h^{(1)}(w),
\]
Otherwise, if $w\leq\hat w_j$, the number tasks of type $j$ are are associated with $\chi_h$. Define 
\[
\chi_h^{(2)}(w)=\frac{w-a\theta_{j-1}}{\theta_j-\theta_{j-1}}, \; \;
\chi_l^{(2)}(w)=a-\chi_h^{(2)}(w),
\]
Observe, that in the first case the buffer is full with tasks of type $j-1,j$, while in the second case the buffer is full with tasks of type $j,j+1$.
The way we treat this is by bounding $\Del^n$ from above by a quantity that depends on
$\eps_1$, rather than by an arbitrarily small $\eps''$. \AddedTh{Recall that $\eps_1$ refers to the bound applied for the $C$-tightness of $X^{\circ,n}$ in~\eqref{39} and is used to set the size of each interval in the covering $\X_k$ and $\tilde\X_k$.}

Define $c_1=4/\tilde\theta_{\rm min}$ and $\tilde\theta_{\rm min}=\min_{i}(\theta_{i+1}-\theta_{i})$, where for convenience, we denote $\theta_0=0$.
We have for any $w\in\tilde\X_k$, $|w-\hat w_j|\le 4\eps_1$, since
$\hat w_j$ is also in $\tilde\X_k$.
Now, if $w\le\hat w_j$,
then for coordinate $j$ of the minimizing curve it holds $\gamma^a_{j}(w)=\chi_h^{(2)}$.
In this case,
\[
w=(a-\chi_l^{(2)})\theta_h+\chi_l^{(2)}\theta_l=\hat w_j-\chi_l^{(2)}(\theta_h-\theta_l)
\]
We have $\chi_l^{(2)}\leq4\eps_1/(\theta_{j}-\theta_{j-1})$, and consequently $\gamma^a_{j-1}(w)\leq4\eps_1/(\theta_{j}-\theta_{j-1})$. Thus, $\gamma^a_{j}(w)=a-\chi_l^{(2)}>a-\frac{4\eps_1}{\theta_{j}-\theta_{j-1}}$.
This shows that on $\Om^{n,k}$,
\begin{equation}\label{64}
\gamma^a_{j}(X^{\#,n}(t))\ge
a-c_1\eps_1,\qquad
t\in\bT^n,
\end{equation}
Otherwise, if $w\ge\hat w_j$, it follows that $\gamma^a_{j-1}(w)=\chi_l^{(1)}$, and
\[
w=\chi_h^{(1)}\theta_h+(a-\chi_h^{(1)})\theta_l=\hat w_j+\chi_h^{(1)}(\theta_h-\theta_l)
\]
We have $\chi_h^{(1)}\leq4\eps_1/(\theta_{j+1}-\theta_{j})$ and consequently $\gamma^a_{j}(w)\leq4\eps_1/(\theta_{j}-\theta_{j-1})$. Thus,
$\gamma^a_{j}(w)>a-\frac{4\eps_1}{\theta_{j+1}-\theta_{j}}$. This shows that in this case on $\Om^{n,k}$,
\begin{equation}\label{64a}
\gamma^a_{j-1}(X^{\#,n}(t))\ge
a-c_1\eps_1,\qquad
t\in\bT^n,
\end{equation}
Consequently, by position of $X^{\#,n}$ associated with the region $\tilde\X_k$ in case (II), we have by~\eqref{64} and~\eqref{64a} that the minimizing curve is bounded away from $X^{\#,n}$ by order of $\epsilon_1$.
Now, \eqref{41} is valid for all $i\notin\{j-1,j,j+1\}$, by the proof given in case (I).

For the case $j^n=2$, if $w\ge\hat w_1$, the proof is identical. In the case $w\le\hat w_1$, by~\eqref{4:lin_std}, $\gamma^a_{1}(w)=w/\theta_1$. Since $\hat w_1=a\theta_1$, it follows $a\theta_1-w\leq4\eps_1$ and $\gamma^a_{2}(w)\geq a-4\eps_1/\theta_1\geq a-c_1\eps_1$. This shows that~\eqref{64} holds in this case as well.
Finally, for the case $j^n=1$, if $w>\hat w_j$ then the proof is similar to the cases $1\leq j^n<J$.
(the fact that $\gamma^a_{1}$  may assume the value zero does not affect the proof).
Note that~\eqref{41} is valid for all $i>2$, by the proof given in case (I).
For the particular case $j^n=J$ we have $\hat w_J=a^*$, because this is the maximal buffer load.
Thus, we have only the case where $w\leq\hat w_J$, which is treated in the similar way.

Combining all the estimates 
for all small $\eps''$,
\[
\PP(\Om^{n,k}\cap\{\max_{i\ne j}\Del^n_i(\zeta^n)>\eps''\})\to0.
\]
The estimate \eqref{52} and
the bounds \eqref{64},\eqref{64a} give
$\PP(\Om^{n,k}\cap\{\Del^n_{j}(\zeta^n)>2c_1\eps_1\})\to0$ as $n\to\iy$.
This gives
\[
\PP(\Om^{n,k}\cap\{\max_{i\le I}|\Del^n_i(\zeta^n)|>4c_1\eps_1\})\to0,
\]
as $n\to\iy$. 

We now determine the constant $c_0$ used to define $K$. \AddedTh{The objective is to determine the relation between $\eps'$, which used to set $\zeta^n$, the first time the difference between the scaled process and the minimizing curve is greater than $\eps'$, and $\eps_1$.}
We do this in such a way that $4c_1\eps_1<\eps'/2$. In particular, any constant $c_0>8c_1\bx
=32\bx/\tilde\theta_{\rm min}$ will do.

We have bounded $\Del^n_i(\zeta^n)$ by arbitrarily small $\eps'$ in case (\textbf{\emph{I}}) and by $\eps_1$ in case (\textbf{\emph{II}}), and set the relation between these two bounds.
This way we obtain $\PP(\Om^{n,k})\to0$ as $n\to\iy$.

\subsubsection*{(\textbf{III})} $0\in\tilde\X_k$.
The is no difference of this case from case (I) with exception that $\hat X^n$ may be at
zero at some $t$. 
The analysis in case (I) results in the same conclusion, namely $\PP(\Om^{n,k})\to0$ as $n\to\iy$.

\subsubsection*{(\textbf{IV})} $a^*\in\tilde\X_k$. This case corresponds to the rejections at free boundary. 
It is treated by adding a negative term in the equations written in case (I). 
(Note that we can pick $\eps$ to be sufficiently small to avoid treating (II) here).

Having shown that $\PP(\Om^{n,k})\to0$ in all cases, using \eqref{63} and the fact that
$\del>0$ is arbitrary completes the proof of the lemma.

As a consequence of the lemma and \eqref{60}, we have $\PP(\sig^n<T)\to0$ as
$n\to\iy$. Since $\eps'$ is arbitrary, \eqref{35} is established.

To accomplish the proof one has to show the weak convergence of the costs, i.e.
\[
\int_0^\iy e^{-\al t}[h\cdot\hat X^n(t)+\al r\cdot \hat Z^n(t)]dt]\To
\int_0^\iy e^{-\al t}[h\cdot\gamma^a(\bar X^a(t))+\al \bar r \bar Z^a(t)]dt,
\]
and to show the uniform integrability of rejection and holding cost components to obtain
\[
\lim_nJ^n(U^n)=\lim_n\mathbb E\Big[\int_0^\iy e^{-\al t}[h\cdot\hat X^n(t)+\al r\cdot\hat Z^n(t)]dt\Big]
=\mathbb E\Big[\int_0^\iy e^{-\al t}[h\cdot\gamma^a(\bar X^a(t))+\bar r\bar Z^a(t)]dt\Big]=V(x_0;\eps)
\]
Then, the final result follows once $V(x_0,\eps)\to V(x_0)$ as $\eps\to0$ is seen. These technical steps are covered in~\cite{Atar-Shif} and are omitted here.
Therefore, this accomplishes the proof of the theorem.

\section{Conclusion}\label{sec:concl}
In this paper, we analyzed the problem of service scheduling and admission of multi-class tasks arriving to the shared buffer. The main motivation for this problem comes from the growing utilization of cloud computing resources.
We demonstrated that in the limit, the queuing control is expressed via one-dimensional reduced workload control problem. 
Using the fact the the value function of the reduced and multi-task problem in the limit are coincide,
we formulated the policy for which we proved, that once applied to the multi-task scaled problem, it attains the upper bound for the value function. Hence, we proved that the AO policy constitutes a form of state-space collapse. In particular, we demonstrated that the shared buffer is either occupied by two classes of tasks, or it is not full.
The classes of the tasks are determined by the workload. 

The workload process notion has a particular importance in computing systems. Namely, it indicates the remained computing effort till the buffer becomes empty. 
The AO policy we presented has an advantage of being simple to implement, which is important  
in cloud computing, then the managing overhead is of sensible weight.  
Moreover, as the rate of incoming tasks and cloud computing capabilities increase, the policy becomes more close to the optimal one and the optimality gap decreases. We conclude that diffusion approximation method has applicative meaning of constantly growing importance in this context. We suggest that other cloud computing configurations and computer networks in general can be similarly treated. 


\bibliographystyle{is-abbrv}

\bibliography{refsSh}







\end{document}